\documentclass[final,smallextended]{svjour3}
\usepackage{graphicx}
\usepackage{mathptmx}
\usepackage{units}
\newcommand{\WICD}{W_\textit{\scriptsize{ICD}}}
\newcommand{\Weq}{W_\textit{\scriptsize{eq}}}

\journalname{Journal of Statistical Physics}
\begin{document}

\title{
Solution of the Fokker-Planck equation with  a logarithmic potential
} 

\author{A. Dechant \and E. Lutz \and E. Barkai \and D. A. Kessler}
\institute{A. Dechant \at Department of Physics, 
University of Augsburg, 
D-86135 Augsburg, Germany \\
\email{andreas.dechant@physik.uni-augsburg.de}
\and E. Lutz \at
Department of Physics,
University of Augsburg,
D-86135 Augsburg, Germany\\
\email{eric.lutz@physik.uni-augsburg.de} 
\and E. Barkai \at
Department of Physics, Bar--Ilan University, Ramat-Gan
52900 Israel \\
\email{eli.barkai@biu.ac.il} 
\and D. A. Kessler \at
Department of Physics, Bar--Ilan University, Ramat-Gan
52900 Israel \\
\email{kessler@dave.ph.biu.ac.il} }


\date{Received:   / Accepted: }
\maketitle

\begin{abstract}
We investigate the  diffusion of  particles in an attractive 
one-dimensional potential that grows logarithmically for large $|x|$ using  the  Fokker-Planck equation.
 An eigenfunction expansion shows that the Boltzmann
equilibrium  density
  does not fully describe  the long time limit of this problem. 
Instead this  limit is characterized by an  infinite covariant density.
This non-normalizable density yields the mean square displacement
of the particles, which  for a certain range of  parameters 
exhibits anomalous diffusion.  In a symmetric potential with an asymmetric initial condition, the average position 
decays anomalously slowly.  This problem also has applications outside the thermal context, as in the diffusion of the momenta
of atoms in optical molasses.

\end{abstract}
%

\section{Introduction}

 Diffusion of particles in the  presence of an attractive logarithmically growing potential 
has attracted  much interest  since it serves as a model for diverse physical
systems. Examples include
charged particles in the  vicinity of a  long  and uniformly
charged polymer\cite{Manning}  (the Manning  condensation problem), diffusive
spreading  of momenta of two-level  atoms in optical lattices\cite{Dalibard,Zoller,Lutz,Renzoni},
single  particle models of long ranged interacting systems\cite{Bouchet,Bouchet1,Chavanis},  probe particles in a 1-d driven fluid\cite{MukamelEPL}, dynamics of
bubbles in  double stranded DNA \cite{Fogedby1,Fogedby,Kafri,Segal},  
vortex dynamics  \cite{Bray}, and nano-particle dynamics in
appropriately constructed  force  fields\cite{Cohen}.  
When  the motion is in contact with a  heat bath, the  steady  state
is given by the equilibrium Boltzmann distribution,
$W_{{\it eq}} (x) \propto  \exp[- V(x)/ k_B T]$.  For a logarithmically growing potential,
$V(x) \sim V_0 \ln(x/a) $ for $x \gg a$,  $V_0>0$, we get an equilibrium distribution with a power-law tail,
$  W_{{\it eq}}(x) \sim x^{ - V_0 / k_B T}$.
 By definition this steady state is normalizable when
 $Z \equiv \int_{-\infty} ^\infty  dx\, \exp[-V(x)/k_B T] $  is finite,
a necessary condition being
$V_0/k_B T > 1$.  
The well investigated Bessel process~\cite{Martin,Schehr} considers a purely logarithmic potential, $V(x) = V_0 \ln|x|$,  which is not regular at the origin.
This  means that the steady state is not integrable, diverging either at large or small $x$, depending on whether $V_0/k_B T$ is smaller or larger than unity. In most cases, the potential is only asymptotically logarithmic, with  a  regular small $x$   behavior characterized by some length scale $a$.
For example,  this length is the  diameter  of the
long  polymer in Manning's problem. 
 Similarly for optical lattices,  the potential (here, in momentum space) is $V(p) = \nicefrac{1}{2}\ln ( 1 + p^2/p_0^2)$, so that $p_0$ sets the length scale.

Here we 
study the case  of diffusion 
in such a regular logarithmic type potential. 
We mainly discuss the case where there exists
a time independent  steady state, 
i.e., the Boltzmann equilibrium distribution is normalizable. 
Nevertheless, higher moments of the distribution are infinite, due to the slow spatial decay of $W_{eq}$.  For example, the second moment, $\langle x^2 \rangle_\textit{eq} = \int_{-\infty}^\infty x^2 W_{eq}(x) dx $,  diverges for $V_0/k_B T < 3$, and the fourth moment for $V_0/k_B T < 5$. The equilibrium distribution is 
not sufficient then to compute all moments higher than some given moment, depending on the size of $V_0$.  Rather, as we will see, the higher moments grow in time, approaching the 
equilibrium value of infinity in the infinite time limit.  At any finite time, however, the moments are perfectly finite,
as the physics demands.  The computation of these moments in the long-time limit involves a new object, which we call the 
 infinite covariant  density. The term infinite in the name~\cite{TK,TK2,TK1} refers to the fact that this density is not normalizable due to its divergence at the origin.  Nevertheless, the
 higher moments  of this distribution are well defined.  The term covariant refers to the fact that the density has a scaling form, which gives rises to the
 power-law growth of the moments with time. 
The scaling properties of the distribution and the power-law growth of the moments have been obtained in Refs.  \cite{MukamelEPL,KesslerPRL} 
 by assuming a  scaling  ansatz.  
Here we construct the time-dependent probability density from first principles via an eigenvalue expansion of the Fokker-Planck operator. 
This new derivation will allow us to calculate additional properties of the system, in particular the anomalous decay of the first moment from an asymmetric initial condition.  This long-term memory is a sign of the breakdown of ergodicity for too small $V_0$~\cite{Gang4}.

The plan of the paper is as follows: We first construct the Schr\"odinger operator associated with the Fokker-Planck equation, and express the probability density function (PDF) in terms of the eigenstates of this operator.  We then specialize to the exactly solvable case of a pure logarithmic potential connected to an inner region of constant potential.  This allows us to explicitly solve for the eigenstates and construct the PDF, and to exhibit its scaling properties in the long-time limit.  In the following section, we generalize the procedure to an arbitrary logarithmic-type potential, which reproduces our previous scaling solution~\cite{KesslerPRL} for the PDF.  We calculate all even moments to leading-order, and the first-order corrections for the zeroth and second moments. In the final section,
we discuss the decay of the first moment.  We conclude with some final observations.

\section{General Formalism} 

We  consider Brownian  particles in a  potential
as  modeled by the Fokker-Planck equation~\cite{Risken}
\begin{equation}
{{\partial} \over {\partial} t} W(x,t) = \hat{L}_{{\rm FP}} W(x,t).
\label{eq01}
\end{equation}
Here $W(x,t)$ is the  probability  density  function (PDF) of finding  a particle 
at $x$ at time $t$ and 
\begin{equation}
\hat{L}_{{\rm FP}} = D \left[ {\partial^2 \over \partial x^2} - {\partial \over \partial x} F(x) \right]. 
\label{eq02}
\end{equation}
subject to some initial density $W_0(x)$ at time $t=0$, where $D$ is the diffusion constant.
If the system is  close to  thermal equilibrium,  the Einstein relation  
gives  $F(x)  =  f(x) /  k_B T $  where $T$ is the temperature,
 and $f(x)$ is the external force acting on the particles.
 For other processes not necessarily  close
to thermal equilibrium, e.g. atoms  in optical lattices~\cite{Dalibard,Zoller},
the  temperature is not defined  in its usual sense, 
and in what follows we do not limit ourselves to thermal systems. 
The  force $F(x)$ is derived from a potential $U(x)$, 
$F(x)= - U'(x)$, which for a thermal system is related to the potential energy $V(x)$ by  $U(x)\equiv V(x)/k_B T$, Then, the normalized steady state solution of the  Fokker-Planck equation  is
given as usual by 
\begin{equation}
W_{{\it eq}} (x) = \frac{1}{Z} e^{ - U(x)}; \qquad Z\equiv \int_{-\infty}^{\infty} dx\, e^{- U(x)}  \ ,
\label{eq03}
\end{equation}
 which for a thermal system  is the Boltzmann distribution.
 
 The formal solution of the Fokker-Planck equation is given in terms
of the eigenfunction expansion~\cite{Risken}
\begin{equation}
W(x,t)  = e^{ -U(x)/2}  \sum_{k=0} ^\infty a_k \psi_k (x) 
e^{ - \lambda_k t} \ ,
\label{eq06a}
\end{equation}
where
\begin{equation}
a_k = \int_{-\infty}^{\infty} dx\, W_0(x) \psi_k(x) e^{U(x)/2} \ .
\end{equation}
In particular, for a $\delta$-function initial condition, $W_0(x)=\delta(x-x_0)$, $a_k = e^{U(x_0)/2} \psi_k(x_0)$.  We will mostly concern ourselves here with the case $x_0=0$, examining the more general situation at the end, in which case 
\begin{equation}
W(x,t)  = e^{ -U(x)/2+U(0)/2}  \sum_{k=0} ^\infty \psi_k(0) \psi_k (x) 
e^{ - \lambda_k t}.
\label{eq06}
\end{equation}
The  wave  functions  $\psi_k(x)$ satisfy  the  Schr\"odinger equation
\begin{equation}
\hat{H}  \psi_k (x) =  - \lambda_k \psi_k (x)  \ ,
\label{schro}
\end{equation}
 where 
the effective Hamiltonian is the similarity transform of the Fokker-Planck operator:
\begin{equation}
\hat{H} = e^{ U(x)/2} \hat{L}_{{\rm FP}} e^{-U(x)/2} = D\left[ \frac{d^2}{dx^2} - \nicefrac{1}{2} F'(x) - \nicefrac{1}{4} F^2(x)\right] \ ,\label{eq08}
\end{equation} 
so that
\begin{equation}
V_\textit{\scriptsize{eff}} (x) = D\left(\nicefrac{1}{2}F'(x) + \nicefrac{1}{4}F^2(x)\right)
\label{veff}
\end{equation}
is the effective potential of the Schr\"odinger problem..

One eigenstate is immediate, namely the $\lambda_k=0$
bound state, which is simply 
\begin{equation}
\psi_0(x)=Z^{-\nicefrac{1}{2}} e^{-U(x)/2}
\end{equation}
so that $a_0=Z^{-1/2}$, independent of the initial state $W_0(x)$.
This corresponds  to the ground state of the effective Hamiltonian (since it has no zeros) and 
 describes
the steady state. Indeed the  $k=0$ term in Eq. 
(\ref{eq06}) is the  steady  state Eq. 
(\ref{eq03}). 
In many situations (e.g., a binding harmonic  potential)
 the  eigenspectrum of the Fokker-Planck equation is discrete and then 
in the long  time  limit the steady state  is the solution of the problem, with corrections that decay exponentially in time. 
In other cases, the spectrum is continuous,  but
then  a steady  state  is not normalizable (e.g. free particle on the infinite line). 

We are interested here in the case of a (reflection symmetric) logarithmic potential, defined by its asymptotic large-$x$ behavior, $U(x)\approx U_0\ln(x/a) + o(1)$. This case is different from  the more typical examples discussed above.  For $U_0>1$
there is a continuous spectrum starting at zero, since $F$, $F'$ go to zero for large $x$ and so the effective potential of the Schr\"odinger equation, given by Eq. (\ref{veff}), vanishes for large $x$. In addition, there is the zero energy ground state, since  a normalizable steady state exists, lying at the edge of the continuum.
 As we  will  demonstrate,
the steady state, and hence the Boltzmann equilibrium distribution for a thermal  system,
does not fully describe the  density profile of the particles at long but finite times in the case of the logarithmic potential with $U_0 > 1$.  For $U_0<1$, on the other hand, there is no normalizable steady state, and the spectrum is pure continuous extending from $E=0^+$..  
For a general logarithmic potential, of course, the continuum states cannot be solved for exactly, so we proceed in the next section by solving a specific simple example.

\section{A Solvable Example}
A simple potential which admits an exact solution is 
\begin{equation}
U(x) = \left\{
\begin{array}{c c} 
0  & |x| < a \\ 
U_0 \ln |x|/a & |x|\ge a 
\end{array}\quad,
\right.
\label{toyU}
\end{equation}
which for the thermal case is associated with the potential energy $V(x)=\theta(|x|-a) k_B T U_0 \ln(|x|/a)$.
The associated $F(x)$ in the Fokker-Planck equation is 
\begin{equation}
F(x)  = \left\{ 
\begin{array}{c c}
0  &  |x| < a \\
-  { U_0 \over  x} & |x|> a
\end{array} 
\right. \quad,
\label{eq04}
\end{equation}
 and the steady-state PDF is 
\begin{equation}
W_{{\rm eq}}(x) =\frac{1}{Z} e^{-U(x)} =  \left\{
\begin{array}{c c}
\frac{1}{Z} & |x| < a \\
\ & \\
\frac{1}{Z} \left(\frac{|x|}{a}\right) ^{-U_0} & |x| \ge a 
\end{array}
\right.\quad,
\label{eq05}
\end{equation}
with
\begin{equation}
Z = {2aU_0 \over U_0 - 1 } \ .
\label{Zsolv}
\end{equation}
The  logarithmic potential yields a density  profile which is a power law.
As noted above, for  the steady  state to exist we must have  $U_0 > 1$, otherwise the steady 
state  is not normalizable.
Notice that  in the  steady state
the second moment 
$\langle x^2  \rangle=\int_{-\infty} ^\infty dx\,  x^2 W_{{\rm eq}} (x)  = \infty$ 
 diverges if $1 < U_0 < 3$. Physically for any finite long time, with the particles starting on the origin, 
the mean square displacement cannot be larger than what is 
permitted by diffusion $\langle x^2 \rangle \le 2 D t$. 
In what  follows,
we  consider  $U_0 > 1$  and initially at time $t=0$ the 
particles are  located at the  origin  $x=0$. 

For our solvable example, Eq. (\ref{toyU}), the Schr\"odinger equation (\ref{schro})  reads
\begin{equation}
D\left[ -  {d^2 \over d x^2} 
+ \theta(|x|-a) \left( {U_0^2  + 2 U_0 \over  4x^2}  \right)   - {\delta(|x|-a)  \over 2a} U_0 \right] \psi_k = \lambda_k \psi_k 
, 
\label{eq09}
\end{equation} 
where  the Heaviside function $\theta(|x|-a)=1$  if  $|x|>a$ and zero otherwise. As mentioned above, we choose for now an initial condition with the particle at the origin, from which it follows, given the symmetry of the potential,  that 
 the solution $W(x,t)$ also has even  parity. Hence in our  case  it is sufficient 
to investigate  only  the even  solutions  of  the Schr\"odinger equation. 
As we discussed in Section II, the spectrum consists of the ground state with zero energy and a continuum that goes down to $\lambda_k=0^+$.  We thus write the energy of the continuum states as $\lambda_k = Dk^2$, $k>0$.

 In the  region  $|x|<a$ the solution is trivial, since  the effective
potential in the Schr\"odinger equation is zero, we have  the  behavior  of  a free
particle,  whose even solution reads
\begin{equation}
\psi_k(x) = N_k \cos  k x \ \ \mbox{for} \ \ |x|<a . 
\label{eq10}
\end{equation}
$N_k$  is  a normalization  constant to be determined later.
For $x>a$, the general solution with $\lambda_k=Dk^2$ is
\begin{equation}
\psi_k(x) = N_k \sqrt{\frac{x}{a}} \Big[A_k J_\alpha( k x ) + B_k J_{-\alpha}(kx)\Big] ;\qquad x>a 
\label{eq12}
\end{equation}
with
\begin{equation}
\alpha= {1 + U_0 \over 2} \ ,
\label{eq12a}
\end{equation}
 and since  the wave  functions are  even  it is sufficient  to
consider  $x>0$.

 The coefficients $A_k$  and  $B_k$ are determined from two boundary conditions
at $|x|=a$. First, the  continuity of  the wave 
functions, Eqs. (\ref{eq10}) and (\ref{eq12}), gives
\begin{equation} 
 A_k J_\alpha(ka) + B_k  J_{-\alpha}(ka) = \cos (ka).
\label{eq13}
\end{equation}
The $\delta$ function in the effective Hamiltonian gives a discontinuity in $\psi_k'$ at $x=a$, so that
\begin{equation}
-  \left[ \left.{d  \psi_k \over d x} \right|_{x=a^{+} } - \left.{d \psi_k \over d x}\right|_{x=a^{-}} \right] - U_0{\psi_k(1) \over 2a} = 0 \ .
\label{eq14}
\end{equation}
Eqs. (\ref{eq13},\ref{eq14}) determine the two unknowns 
$A_k$ and $B_k$, which in  principle gives a formal but rather cumbersome
solution  to the problem. 
Luckily  we are  interested only  in the long  time  
limit of the problem where naively one expects the steady  state  
to describe the system. For that limit,  it is sufficient
to  consider only the small $k \to 0$  behavior, since  for long times
the modes with  finite $k$
decay  like $\exp(-  \lambda_k t)  = \exp(  -  D  k^2 t)$ 
and  hence  are negligible. More formally, the summation 
over the  eigenfunctions
in Eq. (\ref{eq06})
 must be replaced 
with  an integration since  the spectrum is  continuous,  (see Eq. (\ref{eq21}) below)  and hence we may  
consider Eq. (\ref{eq06})  as a  Laplace  transform (where $t$  is  the  Laplace variable)
 because of  the  $\exp(- D k^2 t)$  term. 
Then according to the Tauberian theorems,  the long time behavior corresponds
to the $k\to 0$ limit. 

 In the  small $k$ limit we use Eqs. (\ref{eq13},\ref{eq14}) and the small $z$ limit of $J_\alpha(z)\,$\cite{AbSt},
\begin{equation}
J_\alpha(z)  \approx \frac{\left(\frac{z}{2}\right)^\alpha }{\Gamma(1 + \alpha)} \left(1 - \frac{z^2}{4(1+\alpha)} + {\cal{O}}(z^4)\right) \ ,
\label{bessum}
\end{equation}
to find
\begin{equation}
A_{k} \approx {\cal A}  (ka)^{2- \alpha} ;\qquad\qquad B_{k} \approx {\cal B} (ka)^\alpha,
\label{eq15} 
\end{equation} 
where
\begin{equation}
{\cal A} = - 2^{\alpha-2} \Gamma(\alpha-1)  (2 \alpha-1) ;\qquad\qquad
{\cal B} = 2^{-\alpha} \Gamma(1  - \alpha).
\end{equation} 
In particular, $A_k \gg B_k$ for $\alpha>1$.  

 Now we  determine the normalization $N_k$ from  the condition
\begin{eqnarray}
{1  \over 2} &=& \int_0 ^L [\psi_k(x)]^2 dx\nonumber\\
&=& (N_k)^2\left[\int_0^a dx\,\cos^2 kx  + \int_a^L dx\, (x/a)\left[A_k J_\alpha(kx) + B_k J_{-\alpha}(kx)\right]^2 \right]
\label{eq16}
\end{eqnarray} 
where  $L \to \infty$ is the system size (a parameter which will eventually
cancel out). The second integral grows with $L$, and so only the large $x$ behavior of the integrand is relevant. Using the large $z$ asymptotics,
$J_\alpha(z ) \sim \sqrt{ 2/ (\pi z)} \cos( z - \alpha \pi/2 - \pi/4)$, and the fact that $A_k \gg B_k$, we find in the large $L$ limit
   \begin{equation}
(N_k)^2  \approx { \pi \over  2 {\cal A}^2 L }(ka)^{2 \alpha-3}. 
\label{eq17}
\end{equation}

 Having found the coefficients $A_k$, $B_k$ and  $N_k$ we insert
the eigenfunctions Eq. (\ref{eq12})  in the expansion Eq. (\ref{eq06}).
We separate  the time independent part   (the ground state  $k=0$) from
the contributions  of the continuum states ($k>0$) 
and write
\begin{equation}
W(x,t) = W_{{\it eq}}(x) +  W^{*}(x,t) 
\label{eq18}
\end{equation}
where, for $x>a$,
\begin{eqnarray}
W^{*} (x,t) &\approx&
\left(\frac{x}{a}\right)^{1/2-\alpha} \sum_{k>0} 
{ \pi \over 2 {\cal A}^2 L } (ka)^{ 2\alpha - 3} e^{-Dk^2 t} \times\nonumber \\
&\ & \qquad\qquad\qquad  \sqrt{\frac{x}{a}}\left[ 
{\cal A}  (ka)^{2-\alpha} J_{\alpha} ( k x) + {\cal B} (ka)^{\alpha}  J_{- \alpha} ( k x) \right] \ .
\label{eq19}
\end{eqnarray}
In the large $L$ limit, the sum of $k$ becomes an integral (requiring $\psi_k$ to vanish at $x=L$, for example, implies an asymptotic spacing of $\Delta k = \pi/L$) so that
\begin{equation}
\sum_{k>0} ^\infty \cdots  \to \int_0 ^\infty  {   Ldk \over \pi} \cdots .
\label{eq21}
\end{equation} 
We can now rescale $k$ in the integral by $\sqrt{Dt}$ and investigate the time dependence of $W^*$.  In particular, for $|x|<a$, one finds that $W^*$ decays as $t^{1-\alpha}$, and so is smaller than $\Weq$ in this domain.  For $|x|>a$, the $J_{\alpha}$ term scales as $t^{-\alpha/2}$ whereas the $J_{-\alpha}$ term scales as $t^{1-3\alpha/2}$, which decays faster for $\alpha>1$, i.e. $U_0 > 1$.  Thus, for $x>a$, using Eq. 6.631.5 of Ref. \cite{Grad}
\begin{eqnarray}
W^*(x,t) &\approx&  \frac{1}{2{\cal{A}}} x^{1-\alpha} a^{2a-2} (Dt)^{-\alpha/2} \int_0^\infty dk\, k^{\alpha-1} J_\alpha\left(\frac{kx}{\sqrt{Dt}}\right)e^{-k^2} \nonumber\\
&=& -\frac{1}{\Gamma(\alpha-1)(2\alpha-1)a} \left(\frac{4Dt}{a^2}\right)^{1/2-\alpha} z^{1-2\alpha} \left[\Gamma(\alpha)-\Gamma\left(\alpha,z^2\right)\right]\ .
\end{eqnarray}
Here
\begin{equation}
z  \equiv{ x  \over \sqrt{4Dt} } 
\label{eq20}
\end{equation} 
is the scaling variable, which is immediately familiar as Brownian scaling, and we are considering
 the limit of  large times with $z$ fixed.  $\Gamma(a,x)$ is the incomplete Gamma function\cite{AbSt}. Plugging into Eq. (\ref{eq18}) and rewriting $\Weq$ in terms of $\alpha$  gives
\begin{equation}
  W(x,t) \approx \underbrace{\frac{\alpha-1}{(2\alpha-1)a} \left(\frac{x}{a}\right)^{1-2\alpha}}_{W_\textit{\tiny eq}(x)} + \underbrace{\frac{-1}{\Gamma(\alpha-1)(2\alpha-1)a} \left(\frac{4Dt}{a^2}\right)^{1/2-\alpha} z^{1-2\alpha} \left[\Gamma(\alpha)-\Gamma\left(\alpha,z^2\right)\right]}_{W^{{}^*\!}(x,t)} \ .
\label{eq28}
\end{equation} 
The  first term  on the  right  hand side is the  ground state, namely the usual Boltzmann equilibrium. Crucially, this is exactly cancelled by the first 
 term of $W^*$, which is essential since $W$ is cut off and thus much smaller than $\Weq$ in the large $x$ region. We are then left with%
\begin{equation}
W(x,t)  \approx   \frac{1}{\Gamma(\alpha-1)(2\alpha-1)a} \left(\frac{4Dt}{a^2}\right)^{1/2-\alpha} z^{1-2\alpha} \Gamma\left(\alpha,z^2\right) \ ,
\label{eq30}
\end{equation}
which  clearly differs from the steady state Eq. (\ref{eq05}). For $z$ small, the $\Gamma$ function is near unity, and $W$  reproduces the ($x > a$) equilibrium state.  For $z$ large, however, $\Gamma(\alpha,z^2)$ decays as a Gaussian, 
\begin{equation}
\Gamma(\alpha,z^2) \approx z^{2(\alpha-1)} e^{-z^2} \ ,
\end{equation}
ensuring the finiteness of all the moments.  For large $z$, then,
\begin{equation}
W(x,t) \approx  \frac{1}{\Gamma(\alpha-1)(2\alpha-1)a} \left(\frac{4Dt}{a^2}\right)^{1/2-\alpha} z^{-1} e^{-z^2}  \ .
\end{equation}
The scaling structure of $W$, namely $W(x,t)=t^{1/2-\alpha} {\cal{F}}(z)$ is exactly the starting point of our scaling ansatz presented in Ref. \cite{KesslerPRL}

We have assumed in the above that $\alpha>1$, i.e.,  $U_0 > 1$.  The situation for $\alpha<1$ is markedly different.  Here there is no normalizable equilibrium solution.  Hence, the $k=0$ mode is not a part of the spectrum of the Fokker-Planck operator.  Also, $B_k \gg A_k$ in the small-$k$ limit, and so
\begin{equation}
(N_k)^2 \sim \frac{\pi}{2{\cal{B}}^2L} (ka)^{1-2\alpha}\ .
\end{equation}
  Thus, for $\alpha<1$, we have, for $x>a$, using Eq. 6.631.4 of Ref. \cite{Grad},
\begin{eqnarray}
W(x,t) &\approx&  \frac{1}{2{\cal{B}}a} x^{1-\alpha} a^{5/2-3\alpha} (Dt)^{-1+\alpha/2} \int_0^\infty dk \,k^{1-\alpha} J_{-\alpha}\left(\frac{kx}{\sqrt{Dt}}\right)e^{-k^2} \nonumber\\
&=& \frac{1}{\Gamma(1-\alpha)a} z^{1-2\alpha} \left(\frac{4Dt}{a^2}\right)^{-1/2}e^{-z^2} \ .
\label{smallalfbigx}
\end{eqnarray}
 For fixed $x$, so that $z \ll 1$, it behaves as $t^{-1+\alpha} x^{1-2\alpha}$.  In other words, the central region behaves like the
nonnormalizable equilibrium solution, with a prefactor that decays in time.  We also see this by looking at the behavior for $x<a$, where
\begin{eqnarray}
W(x,t) &\approx& \frac{1}{2{\cal{B}}^2 a} (Dt)^{\alpha-1} \int_0^\infty dk\, k^{1-2\alpha} e^{-k^2} \nonumber\\
&=& \frac{1}{\Gamma(1-\alpha) a} \left(\frac{4Dt}{a^2}\right)^{\alpha-1} \ ,
\end{eqnarray}
which, since $\Weq(x)$ is constant for ${x}<a$, is again the equilibrium solution with a prefactor that decays in time.

\section{General Logarithmic Potential}

We now turn to the  general case of a logarithmically growing potential, which will be seen to have the same basic structure as the above calculation.
 Consider a symmetric potential $U(x)$ in such a way that as $|x| \to \infty$ we have
$U(x)  \sim U_0\ln(|x|/a) + o(1)$. We set $U(0)=0$, since a constant shift in the potential can be accommodated by a shift in $a$. We also assume a steady state exists,  that is the potential is sufficiently regular at the origin that $Z$ is finite 
and so $\Weq (x) \sim Z^{-1} (x/a)^{ - U_0} $
 for $x \to \infty$.
 The partition function $Z$ depends
on the exact shape of the potential, and it is found from  the 
 normalization condition  of the steady state:
 $Z = 2 \int_0^\infty e^{-U(x)}dx$.
 
 Again the key to the calculation is  finding the continuum eigenstates for small $k$.  For large $x$, we can approximate the
 potential by its asymptotic form $U(x) \approx U_0 \ln(x/a)$, and so the Schr\"odinger equation is the same as we solved above, giving
 \begin{equation}
 \psi_k(x) = N_k \sqrt{\frac{x}{a}} \left[ A_k J_\alpha(kx) + B_k J_{-\alpha}(kx) \right] \ ,
 \label{eqIVa}
 \end{equation}
 where again $\alpha = (1 + U_0)/2$. This approximation clearly fails for the central region where $x$ is of order $a$.  Instead, we will solve the Schr\"odinger Eq. as a power-series in $k^2$.  This series breaks down, as we will see, at large $x$.  Nevertheless there is an overlap region $a \ll x \ll 1/k$ where both approximations are valid and will enable us to determine $A_k$ and $B_k$.  To leading order we can drop the $Dk^2 \psi_k$ term in the Schr\"odinger equation, so that $\hat{H}\psi_k \approx 0$, and we have that $\psi_k$ is none other than the zero-energy solution $\psi_0$, up to normalization,
 \begin{equation}
 \psi_k \approx N_k e^{-U(x)/2} \ .
 \end{equation}
 For large $x$, this behaves as $\psi_k \approx N_k (x/a)^{1/2-\alpha}$, which comparing to Eq. (\ref{eqIVa}) for small $kx$ gives us, using the small argument expansion of $J_{\pm \alpha}$, Eq. (\ref{bessum}),
 \begin{equation}
 B_k \approx  {\cal{B}}(ka)^{\alpha}; \qquad \qquad {\cal{B}} = 2^{-\alpha}  \Gamma(1-\alpha) \ .
 \end{equation}
 This agrees with what we found in our special case in the previous section.  To determine $A_k$, however,  we need to go to next order in perturbation theory in $k^2$.

Writing 
\begin{equation}
\psi_k(x) = \psi_{k,0}(x) + (ka)^2 \psi_{k,1}(x) + \ldots \ ,
\label{psipert}
\end{equation}
where $\psi_{k,0} (x)\equiv N_k e^{-U(x)/2}$ is the zeroth order solution.  Inserting into the Schr\"odinger Eq. (\ref{schro}), we find that $\psi_{k,1}$ satisfies the inhomogeneous equation
 \begin{equation}
  \hat H \psi_{k,1} = \frac{D}{a^2} \psi_{k,0} \ .
  \end{equation}
  We can solve for $\psi_{k,1}$ in terms of  two independent solutions to the homogeneous equation, which we take as the even solution $f(x)\equiv e^{-U(x)/2}$ and the odd solution $g(x)$, given by
\begin{equation}
g(x)=f(x)\int_0^x \frac{dy}{af^2(y)} \ ,
\label{eq41}
\end{equation}
with the Wronskian 
\begin{equation}
\textit{Wr}[f,g]\equiv fg'-f' g = 1/a\ .
\end{equation}
It is easily verified that $g$ also solves the homogeneous equation: $\hat{H}g=0$.

We can now construct the  even solution for the first-order correction:
\begin{eqnarray}
\psi_{k,1}(x)&=&\int_0^x ds\, \frac{f(x)g(s)-f(s)g(x)}{\textit{Wr}[f,g]} \frac{D}{a^2} \psi_{k,0}(s) \nonumber\\
&=& \frac{N_k}{a}\left[f(x)\int_0^x g(s)f(s)ds - g(x)\int_0^x f^2(s)ds\right] \ .
\end{eqnarray}
In principle we could add to this an arbitrary multiple of the even homogeneous solution, $f(x)$, but this just amounts to changing the normalization of the zeroth order solution.
The even solution to the finite energy problem to first order in $k^2$ is then  $\psi_k=N_k \Psi_k$ where
\begin{equation}
\Psi_k(x) = f(x) + k^2a \left[f(x)\int_0^x g(s)f(s)ds - g(x)\int_0^x f^2(s)ds\right]\ .
\label{Psi}
\end{equation}

 For large $x$,  
 \begin{equation}
 f(x) \approx (x/a)^{1/2 - \alpha} ;  \qquad x \gg a \ .
 \label{fbigx}
 \end{equation}
 From this follows the large $x$ behavior of $g(x)$, since  the integrand in Eq. (\ref{eq41}) is growing with $y$ and the integral is dominated by the large-$y$ contribution, giving
\begin{equation}
g(x) \approx  \frac{(x/a)^{\alpha+1/2}}{2\alpha} ;  \qquad x \gg a \ .
\label{gbigx}
\end{equation}

\subsection{$1<\alpha <2$ ($1<U_0 <3$)}
As this point, we have to distinguish between the cases $\alpha>1$ and $\alpha<1$, as we did with our solvable model.  We first consider $3>\alpha>1$, with the upper bound ensuring that the second moment of the equilibrium solution diverges. The calculation following exactly along the lines of the $\alpha>1$ calculation of the previous section.
From Eq. (\ref{Psi}), using the large $x$ asymptotics of $f$ and $g$, Eqs. (\ref{fbigx}) and (\ref{gbigx}), we have for the large $x$ behavior of $\Psi_k$:
\begin{eqnarray}
\Psi_k(x) &\approx&  \left(\frac{x}{a}\right)^{1/2-\alpha} + k^2 a\left[\left(\frac{x}{a}\right)^{1/2-\alpha} \int_0^x  \frac{s}{2\alpha a} ds - 
\frac{(x/a)^{\alpha+1/2}}{2\alpha } \left( \int_0^\infty f^2(s)ds  - \int_x^\infty   \left(\frac{s}{a}\right)^{1-2\alpha}ds\right)\right] \nonumber\\
&\approx&   \left(\frac{x}{a}\right)^{1/2-\alpha} + (ka)^2 \left[ \frac{(x/a)^{5/2 - \alpha}}{4\alpha} + \frac{(x/a)^{5/2-\alpha}}{4(\alpha-1)} - \frac{(x/a)^{\alpha+1/2}}{2\alpha} \frac{Z}{2a} \right] \ .
\label{psibigx}
\end{eqnarray}
The dominant correction term for large $x$ scales as $(ka)^2 (x/a)^{\alpha+1/2}$, since $\alpha<3$.  Comparing this to the first, leading order, term, we see that the zeroth order term is dominant as long as  $x \ll k^{-1/\alpha} a^{1-1/\alpha}$ which for $\alpha>1$ is a more restrictive condition than $x \ll 1/k$.  Nevertheless, there is still an overlap region between the two approximations, Eqs. (\ref{eqIVa}) and (\ref{psibigx}) where we can perform the matching and find the coefficients $A_k$ and $B_k$. The first two correction terms in Eq. (\ref{psibigx}) just give ${\cal{O}}(k^2)$ corrections to $B_k$.  The last term however determines $A_k$, giving
\begin{equation}
A_k = {\cal{A}} (ka)^{-\alpha}; \qquad\qquad{\cal{A}} = - \frac{ 2^{\alpha-2} \Gamma(\alpha) Z}{a}   \ .
\end{equation}
Using our value of $Z$ for our solvable model of section III, Eq, (\ref{Zsolv}), we again find agreement with our previous results.

The normalization $N_k$ is the same as before, Eq. (\ref{eq17}), and we get for $x \gg a$,
\begin{eqnarray}
W(x,t)&\approx&\Weq(x) + \frac{\sqrt{x/a}}{2{\cal{A}}} \,e^{-U(x)/2}\int_0^\infty dk\, (ka)^{\alpha-1} J_\alpha(kx)e^{-Dk^2 t}\nonumber\\
&=& \Weq(x) -   \frac{e^{-U(x)/2}}{Z} \left(\frac{x}{a}\right)^{1/2-\alpha}  \left(1 - \frac{\Gamma(\alpha,z^2)}{\Gamma(\alpha)}\right) \ .
\label{allx}
\end{eqnarray}
For large $x$, $\Weq(x) \approx  (x/a)^{1-2\beta}/Z$, and so
\begin{equation}
W(x,t) \approx \frac{ (x/a)^{1-2\alpha}}{Z\, \Gamma(\alpha)} \Gamma(\alpha,z^2) \approx \Weq (x) \frac{\Gamma(\alpha,z^2)}{\Gamma(\alpha)} \ .
\end{equation}
This is exactly what we found using our scaling ansatz, 
\begin{equation}
W(x,t) = \frac{1}{a}(4Dt/a^2)^{\nicefrac{1}{2}-\alpha} {\cal{F}}(z)
\end{equation}
 in our previous work~\cite{KesslerPRL}, which represents the large $x$, large $t$ behavior with 
\begin{equation}
{\cal{F}}(z)=\ \frac{a}{Z\, \Gamma(\alpha)} z^{1-2\alpha}\Gamma(\alpha,z^2) \ .
\end{equation}
This leads us to introduce the concept of the {\em Infinite Covariant Density}, which we define as $\WICD(z,t) = W(x,t) dx/dz$, giving
\begin{equation}
\WICD(z,t) \approx  (4Dt/a^2)^{1-\alpha} {\cal{F}}(z) \ .
\label{scaling}
\end{equation}
$\WICD$, as we shall presently see, allows us to compute, to leading order, the expectation values of the  higher moments of $z$.  In fact, $\WICD$ is the probability density function for $z$, except in a small region near $z=0$, where scaling breaks down at any finite time.
 $\WICD$  plays a dual role to $\Weq$ when considering the moments (at large $t$).  The low-order moments of $\WICD$ diverge, (including the zeroth moment, rendering $\WICD$ non-normalizable), since $\WICD \sim z^{1-2\alpha}$ as $z$ approaches 0, but the corresponding expectation values are given correctly by the moments of $\Weq$.  Alternatively, the high-order moments of $\Weq$ diverge due to its algebraic decay with $x$, while the corresponding expectation values grow as a power-law in time and are given correctly, to leading order, by the high-order moments of $\WICD$, which {\em are} finite. (The small $x$ contribution, which is not correctly described by $\WICD$, makes an ${\cal{O}}(1)$ contribution, which is negligible to leading-order).  Which moments are given correctly by $\WICD$ and which by $\Weq$ is determined by $U_0$ (i.e., $\alpha$).  The expectation value $\langle |x|^q \rangle$ is given by the $q$th moment of $\Weq(x)$ for $U_0 > 1+q$, ($\alpha > 1+q/2$) and
by the $q$th moment of $\WICD$ for $U_0 < 1+q$, ($\alpha < 1+q/2$):
\begin{equation}
\langle |x|^q \rangle \approx \left\{ \begin{array}{ll}2\int_0^\infty dx\, x^q \Weq(x)  \qquad\qquad &  U_0>1+q \nonumber \\
(4Dt)^{q/2} 2\int_0^\infty dz\, z^q  W_\textit{\tiny ICD}(z,t) = a^{q+1} \left(\frac{4Dt}{a^2}\right)^{1+q/2-\alpha} \frac{\Gamma(1 + q/2)}{(1+q/2-\alpha)\Gamma(\alpha) Z} \quad& U_0 < 1+q \end{array} \right. \ .
\label{xsqlead}
\end{equation}
We will presently obtain the correction terms to this equation for the case $q=2$.  It is important to note that Eq. (\ref{xsqlead}) implies that our system for this range of parameters exhibits the widely studied phenomenon of anomalous diffusion~\cite{BK,BK1}.

For $x$'s of order 1, $t \gg 1$, so that $z \ll 1$, we have from Eq. (\ref{allx}),
\begin{eqnarray}
W(x,t) &\approx& \Weq(x) + \frac{e^{-U(x)/2}}{2{\cal{A}}^2} \int_0^\infty dk\, (ka)^{2\alpha-3} f(x) e^{-Dk^2 t} \nonumber \\
&=& \Weq(x) \left [ 1 + \frac{ a(4Dt/a^2)^{1-\alpha} }{Z\,(\alpha-1)\Gamma(\alpha)} \right ] \ .
\label{central}
\end{eqnarray}
Thus, in the central region, the solution is given to leading order by the equilibrium distribution, with a correction which decays algebraically  in time.  This implies in turn that a uniform approximation to $W$ is simply
\begin{equation}
W_\textit{\scriptsize{unif}}(x,t) \approx \Weq(x) \frac{\Gamma(\alpha,z^2)}{\Gamma(\alpha)} \ .
\label{unif0}
\end{equation}
since the ratio of $\Gamma$'s approaches unity as $z\to 0$.   The $\Gamma$ function cuts off the equilibrium distribution as $z$ becomes of order unity. The physics is that the long tail of the equilibrium distribution takes time to establish itself, so that beyond the diffusion scale of $\sqrt{Dt}$, it has a Gaussian cutoff. (For similar cases where tails take time to establish themselves, and so are cut off at any finite time at some large $x$, see~\cite{Ner,Schiff}.)   

\subsubsection{First-Order Correction Terms}
Turning now to the correction term in Eq. (\ref{central}), it is interesting to note that the form of this term could be anticipated, due to the fact that the uniform solution displayed above is not normalized; the presence of the cutoff removes some of the probability from the tail of the otherwise normalized $\Weq$:
\begin{eqnarray}
\int_{-\infty}^{\infty} W(x,t) dx &\approx& 2\int_0^\Lambda \Weq(x) dx + 2\int_\Lambda^\infty \frac{1}{Z} \left(\frac{x}{a}\right)^{1-2\alpha} \frac{\Gamma(\alpha,z^2)}{\Gamma(\alpha)} dx \nonumber\\
&\approx& 1 - \frac{a}{Z(\alpha-1)} \left(\frac{\Lambda}{a}\right)^{2-2\alpha} + \frac{a}{Z\Gamma(\alpha)(1-\alpha)} \left\{\left[ \Gamma(\alpha,z^2)\left(\frac{x}{a}\right)^{2-2\alpha}\right]_\Lambda^\infty + \int_\Lambda^\infty \left(\frac{x}{a}\right)^{2-2\alpha}\frac{x}{2Dt} z^{2(\alpha-1)} e^{-z^2}dx\right\} \nonumber\\
&\approx& 1 - \frac{a}{Z(\alpha-1)} \left(\frac{\Lambda}{a}\right)^{2-2\alpha} + \frac{a}{Z\Gamma(\alpha)(1-\alpha)} \left\{-\Gamma(\alpha)\left(\frac{\Lambda}{a}\right)^{2-2\alpha} + \left(\frac{4Dt}{a^2}\right)^{1-\alpha} \right\} \nonumber\\
&\approx& 1 - \frac{a}{Z\Gamma(\alpha)(\alpha-1)}\left(\frac{4Dt}{a^2}\right)^{1-\alpha} \ .
\end{eqnarray}
In the above, we broke up the integral into two pieces at $x=\Lambda$, with $a \ll \Lambda \ll \sqrt{4Dt}$.  If our goal is restore normalization, we can accoomplish this by  rescaling $W$ by an appropriate time-dependent factor which approaches unity as $t \to \infty$:
\begin{equation}
W(x,t) \approx \frac{Z}{Z(t)}\Weq(x) \frac{\Gamma(\alpha,z^2)}{\Gamma(\alpha)}; \qquad\qquad Z(t)=Z/\left [ 1 + \frac{ a(4Dt/a^2)^{1-\alpha} }{Z\,(\alpha-1)\Gamma(\alpha)} \right ]    \ ,
\label{unif}
\end{equation}
and in particular, $W(0,t) = 1/Z(t)$.

In effect, we have added a correction term to $W(x,t)$.  This new term satisfies the Fokker-Planck equation in the central region to leading order in $1/t$, since the only violation is of  order $1/t$ relative to the correction.  This is agrees of course with our explicit calculation in the central region, where we found precisely this additional term, Eq. (\ref{central}). However, in the scaling region, $z \sim {\cal{O}}(1)$, the correction term does {\em not} satisfy the Fokker-Planck equation at all. Rather, we have to solve for the correct scaling form of the correction, and match it to the correction in the central region.  As this will be important for us when discussing $\langle x^2 \rangle$, we go through the exercise.  The key is to write $W=W_{\textit{\scriptsize{unif}}} + W_1$, where $W_{\textit{\scriptsize{unif}}}$ is as given in Eq. (\ref{unif0}), and assume that the correction $W_1$ satisfies
\begin{equation}
W_1(x,t) \sim (4Dt/a^2)^{3/2-2\alpha} {\cal{F}}_1(z) ; \qquad \qquad   {\cal{F}}_1(z) \sim z^{1-2\alpha} \quad \textrm{for} \quad z \ll 1 \ ,
\end{equation}
where the choice of the time exponent  and the small $z$ behavior is necessary for the matching to Eq. (\ref{central})
~\footnote{There are of course other corrections to $W(x,t)$ arising from the deviations of the potential from a pure logarithm.  If $U(x) \sim U_0 \ln(x) + {\cal{O}}(x^{-\epsilon})$ as $x \to \infty$, there is a correction of relative order $t^{-\epsilon/2}$ to the PDF in the scaling region.  This is subdominant over the
correction we are considering as long as $\epsilon >  2\alpha - 2$.  In particular, for the calculation of $\langle x^2 \rangle$ which is our prime concern, the corrections to $W$ dominate over the central region contribution as long as $\alpha<3/2$, as discussed later, and the correction we consider is thus the dominant one in this regime as long as $\epsilon > 1$.}.  
Plugging this into the Fokker-Planck equation gives
\begin{equation}
{\cal{F}}_1'' + \left(\frac{2\alpha-1}{z} + 2z\right) {\cal{F}}_1' - \left(\frac{2\alpha-1}{z^2} + 6-8\alpha\right) {\cal{F}}_1 = 0 \ .
\end{equation}
The solution which matches onto the central region correction is
\begin{equation}
{\cal{F}}_1(z) = B z e^{-z^2} U(2-\alpha,1+\alpha,z^2); \qquad\qquad B=\frac{a\Gamma(2-\alpha)}{ (\alpha-1)\Gamma^2(\alpha) Z^2} \ ,
\end{equation}
where $U(a,b,x)$ is the Kummer confluent hypergeometric function~\cite{AbSt}.  To perform the matching, we used the small $x$ limit of $U(a,b,x)$, namely
\begin{equation}
U(a,b,x) \approx \frac{\Gamma(b-1)}{\Gamma(a)x^{b-1}}  \qquad\qquad x \ll 1 \ .
\end{equation}
The uniform solution correct to second order is then
\begin{equation}
W(x,t) \approx \Weq(x) \left[ \frac{\Gamma(\alpha,z^2)}{\Gamma(a)} + BZ\left(\frac{4Dt}{a^2}\right)^{1-\alpha}z^{\alpha+1/2}e^{-z^2}U(2-\alpha,1+\alpha,z^2)\right] \ ,
\end{equation}
which exhibits the correct behavior in the scaling region, as opposed to  Eq. (\ref{unif}).

It is important to note that, redoing the calculation for arbitrary $x_0$, the result is unchanged (the contribution of the odd eigenfunctions, while no longer zero, can be shown not to contribute to leading order; this contribution is calculated in the following section).  Thus, the solution we have found is independent of initial conditions, as long as the initial conditions have compact support.  This is actually too strict, and a fast enough decay of the initial condition is sufficient to ensure universality.  However, if the initial condition exhibits a slow enough decay, then there are are always contributions from the region $x>\sqrt{4Dt}$, no matter how large $t$ is, and so universality is broken.  For a discussion of some of the novel effects this can induce, see Ref. \cite{mukamel2}.

\begin{figure}
\includegraphics[width=0.7\textwidth]{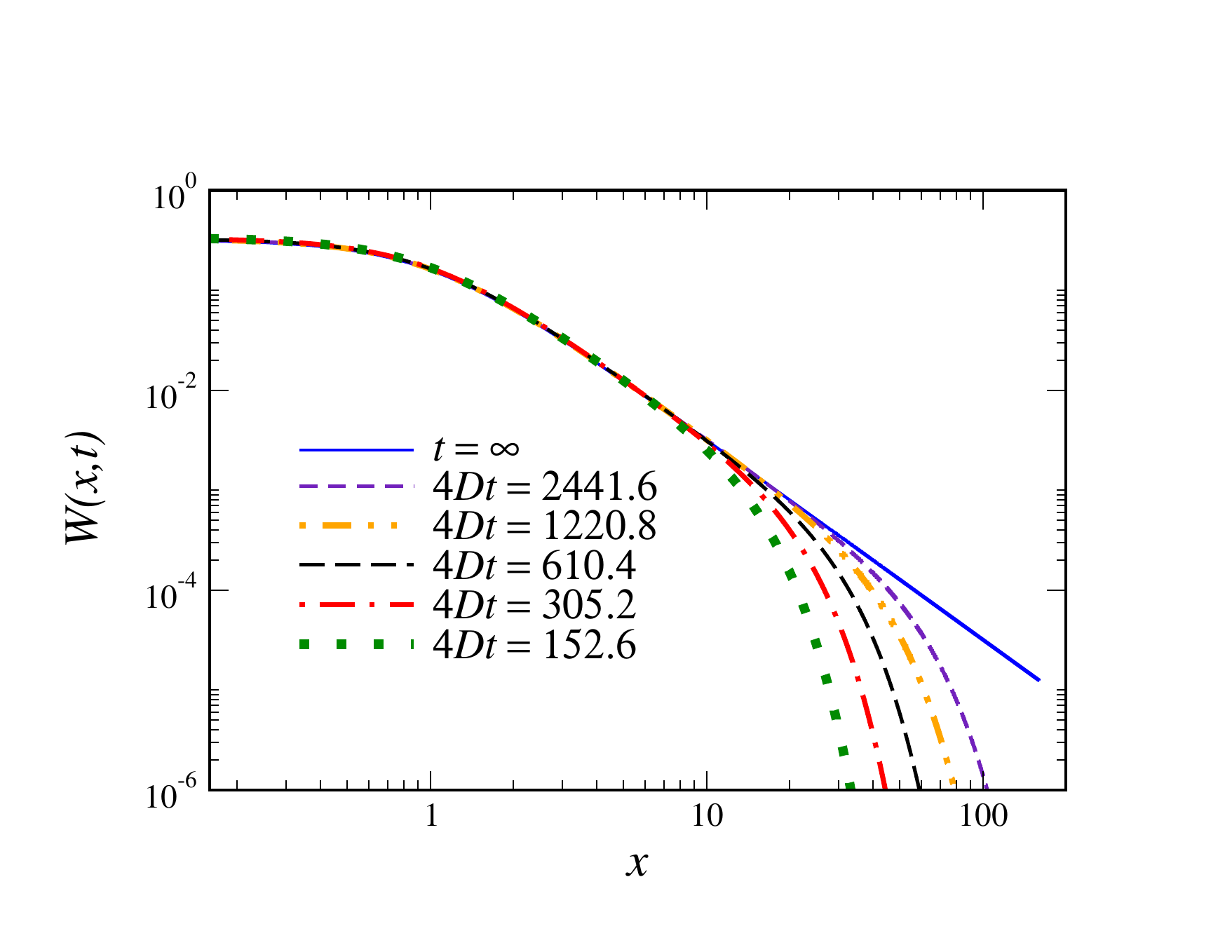}\\
\includegraphics[width=0.7\textwidth]{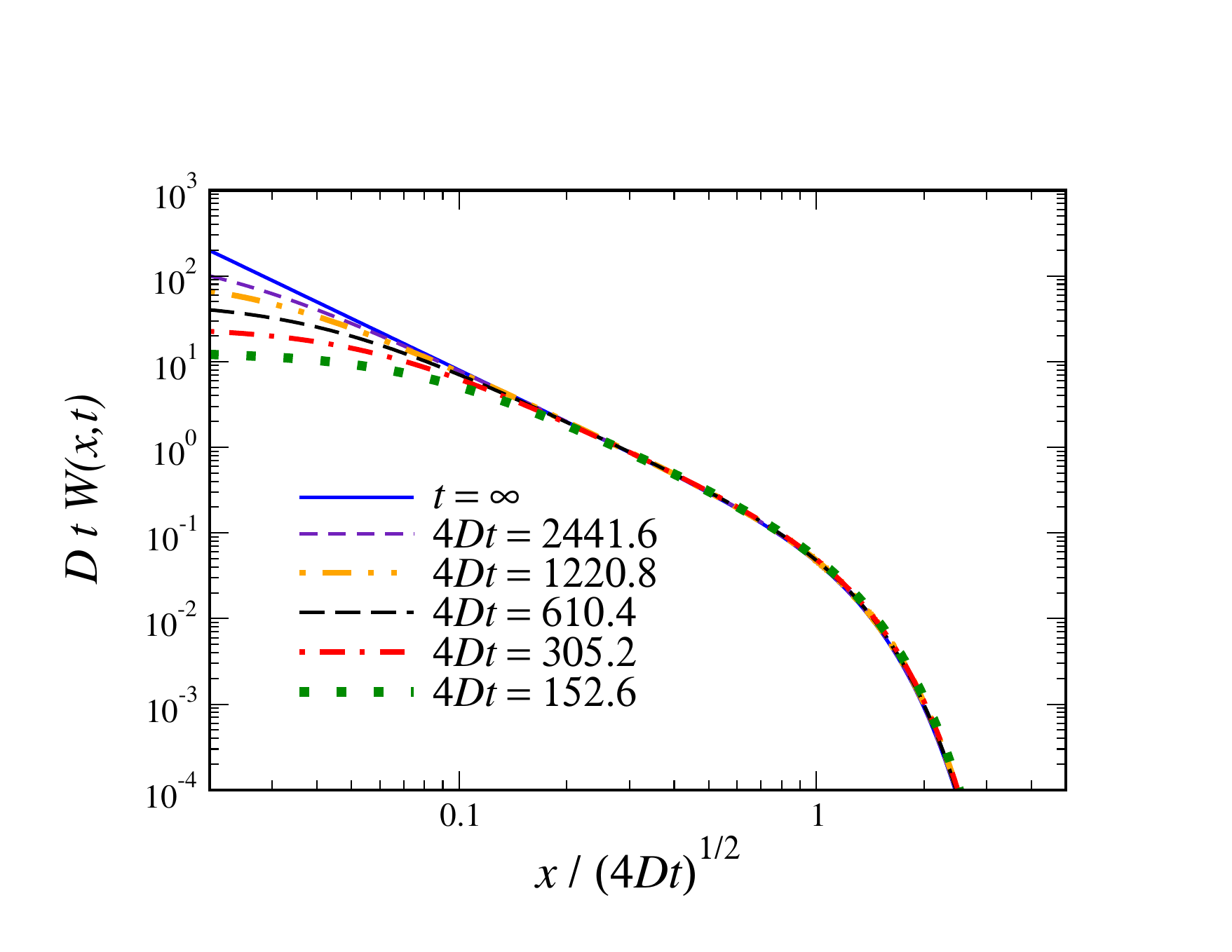}
\caption{Top panel: $W(x,t)$ for various times for the potential $U(x)=\ln(1+x^2)$, ($U_0=2$, $\alpha=3/2$, $Z=\pi$) as calculated by direct integration of the Fokker-Planck equation.  The convergence of the central region to the equilibrium distribution, $\Weq(x)$,
denoted by the line $t=\infty$ is clear, along with the Gaussian cutoff for  $x$ larger than some threshold that moves to larger values with $t$.  Bottom panel:
The scaling collapse $t^{-1/2+\alpha} W(x,t)$ as a function of $x/\sqrt{4Dt}$ showing the approach to the infinite covariant density, $\WICD$, as given in Eq. (\ref{scaling}), denoted by the line $t=\infty$.  The breakdown of the scaling for $x$'s of order 1, which constitutes an ever-shrinking portion of the graph is apparent.}
\label{fig1}
\end{figure}

\begin{figure}
\includegraphics[width=0.7\textwidth]{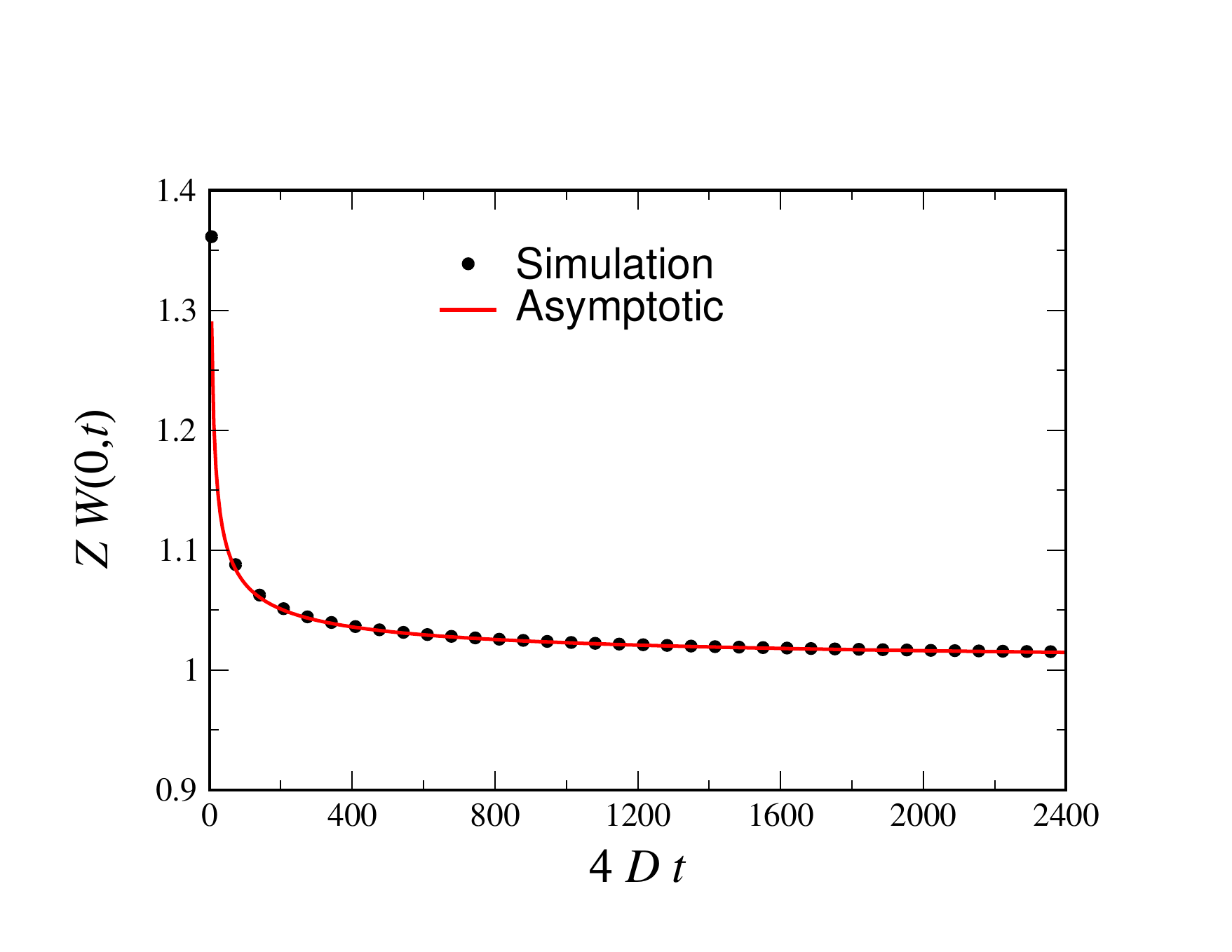}
\caption{$Z W(0,t)$ as a function of time for the case depicted in Fig. 1, together with the long-time analytic prediction, $Z/Z(t)$, as given in Eq. (\ref{unif}).}
\label{fig2}
\end{figure}

In the top panel of Fig. \ref{fig1}, we show $W(x,t)$ for various times for the case of the potential $U(x)=\ln(1+x^2)$, so that $U_0=2$, $a=1$, $\alpha=3/2$.  Together with this we show the equilibrium distribution, marked by $t=\infty$.  The cutoff of the distribution at a point which grows with time is apparent.  The scaling collapse,
where we plot $t^{\alpha-\nicefrac{1}{2}}W(x,t)$ as a function of $x/\sqrt{4Dt}$ is shown in the bottom panel, along with the infinite covariant density, $\WICD$, as given in Eq. (\ref{scaling}), denoted by $t=\infty$. As time increases, the solution in the scaled coordinate approaches  $\WICD$. For any finite long time $t$, expected deviations (which are characterized via our uniform approximation Eq. (\ref{unif0})) from the infinite covariant solution are found for small values of $z$. These deviations become small at $t \gg 1$; however they are important since they indicate that the pathological divergence of the scaling form, Eq. (\ref{scaling}), at the origin is slowly approached but never actually reached; namely, the solution is of course normalizable for finite measurement times.
In Fig. \ref{fig2}, we show $W(0,t)$ together with the theoretical long-time prediction $Z(t)$.

As noted above, the leading-order result for $\langle x^2 \rangle$ is determined by $\WICD$.  When comparing to numerics, one needs to take into account the highest-order correction term, unless the measurement time $t$ is extremely long.  There are two main corrections to the leading-order result. The first
is a result of the correction term discussed above, which leads to a relative correction of order $t^{1-\alpha}$.  The second is the result of the finite contributions for the central region, which have a relative weight of $t^{\alpha-2}$.  Thus, for $\alpha<3/2$, the first is dominant, so that
\begin{eqnarray}
\langle x^2\rangle  &\approx&  \left(\frac{4Dt}{a^2}\right)^{2-\alpha} \frac{a^3}{(2-\alpha)\Gamma(\alpha) Z} + 2\left(\frac{4Dt}{a^2}\right)^{3/2-2\alpha}\int_0^\infty
dx\, x^2 B z e^{-z^2} U(2-\alpha,1+\alpha,z^2) \nonumber\\
&\approx& \left(\frac{4Dt}{a^2}\right)^{2-\alpha} \frac{a^3}{(2-\alpha)\Gamma(\alpha) Z} + \left(\frac{4Dt}{a^2}\right)^{3-2\alpha}Ba^3 \int_0^\infty ds\,  s e^{-s} U(2-\alpha,1+\alpha,s) \nonumber\\
&\approx& \left(\frac{4Dt}{a^2}\right)^{2-\alpha} \frac{a^3}{(2-\alpha)\Gamma(\alpha) Z} + \left(\frac{4Dt}{a^2}\right)^{3-2\alpha}Ba^3 \frac{\Gamma(2-\alpha)}{
\Gamma(4-2\alpha)} \nonumber\\
&=&\left(\frac{4Dt}{a^2}\right)^{2-\alpha} \frac{a^3}{(2-\alpha)\Gamma(\alpha) Z} + \left(\frac{4Dt}{a^2}\right)^{3-2\alpha} \frac{a^4\Gamma^2(2-\alpha)}{(\alpha-1)\Gamma^2(\alpha)\Gamma(4-2\alpha)Z^2} \ .
\label{xsqsmallalf}
\end{eqnarray}
where we have used Eq. 7.621.6 of Ref. \cite{Grad} to integrate the Kummer function.

On the other hand, for $\nicefrac{3}{2}<\alpha<2$ the  leading correction comes from the difference between $\Weq$ and the scaling solution in the central region: 
\begin{equation}
\langle x^2 \rangle \approx a^{3} \left(\frac{4Dt}{a^2}\right)^{2-\alpha} \frac{1}{(2-\alpha)\Gamma(\alpha) Z} - 2\int_0^\infty
dx\, x^2 \left(\frac{1}{Z}\left(\frac{x}{a}\right)^{1-2\alpha} - \Weq(x)\right) \ .
\end{equation}
For example, for the case $U(x)=U_0/2 \ln(1+x^2)$, $U_0 > 2$, we have $Z=\sqrt{\pi}\Gamma(\alpha-1)/\Gamma(\alpha-1/2)$, and
\begin{eqnarray}
\langle x^2 \rangle &\approx&  a^{3} \left(\frac{4Dt}{a^2}\right)^{2-\alpha} \frac{1}{(2-\alpha)\Gamma(\alpha) Z} - \frac{2}{Z}\int_0^\infty
dx\, x^2 \left(x^{1-2\alpha} - (1+x^2)^{1/2-\alpha} \right) \nonumber \\
&\approx& a^{3} \left(\frac{4Dt}{a^2}\right)^{2-\alpha} \frac{1}{(2-\alpha)\Gamma(\alpha) Z} - \frac{2\alpha-1}{Z}\int_0^\infty
dx\, x^2 \int_0^1 ds\,  (s+x^2)^{-1/2-\alpha}  \nonumber \\
&\approx& a^{3} \left(\frac{4Dt}{a^2}\right)^{2-\alpha} \frac{1}{(2-\alpha)\Gamma(\alpha) Z} - \frac{2\alpha-1}{Z}\int_0^1 ds \, s^{1-\alpha} \frac{\sqrt{\pi}\Gamma(\alpha-1)}{4\Gamma(\alpha+1/2)} \nonumber\\
&\approx& a^{3} \left(\frac{4Dt}{a^2}\right)^{2-\alpha} \frac{1}{(2-\alpha)\Gamma(\alpha) Z} -  \frac{\sqrt{\pi}\Gamma(\alpha-1)}{2(2-\alpha)Z\Gamma(\alpha-1/2)} \nonumber\\
&\approx& a^{3} \left(\frac{4Dt}{a^2}\right)^{2-\alpha} \frac{1}{(2-\alpha)\Gamma(\alpha) Z} - \frac{1}{2(2-\alpha)} \ .
\label{xsqbigalf}
\end{eqnarray}

\subsubsection{Numerical Tests}

\begin{figure}
\includegraphics[width=0.4\textwidth]{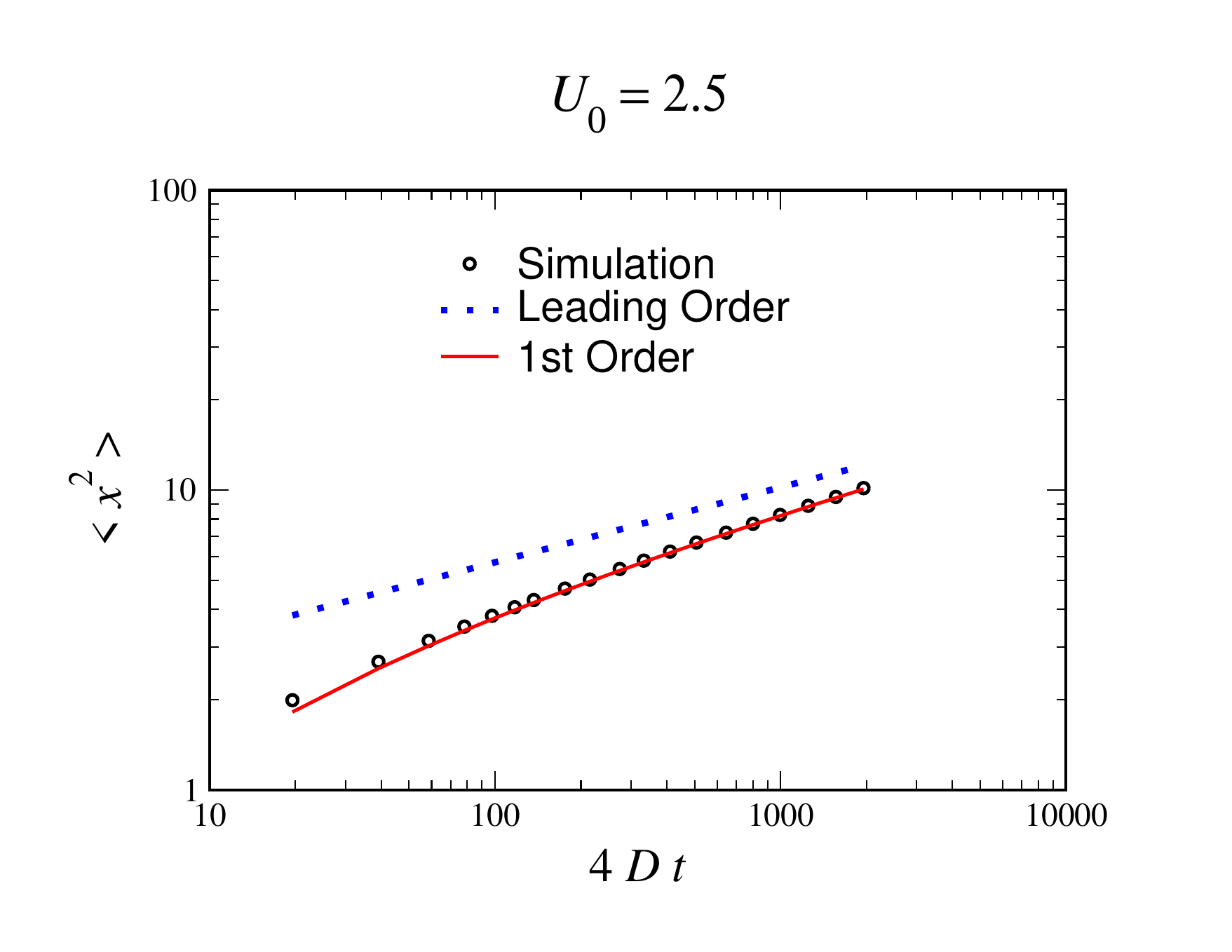}\includegraphics[width=0.4\textwidth]{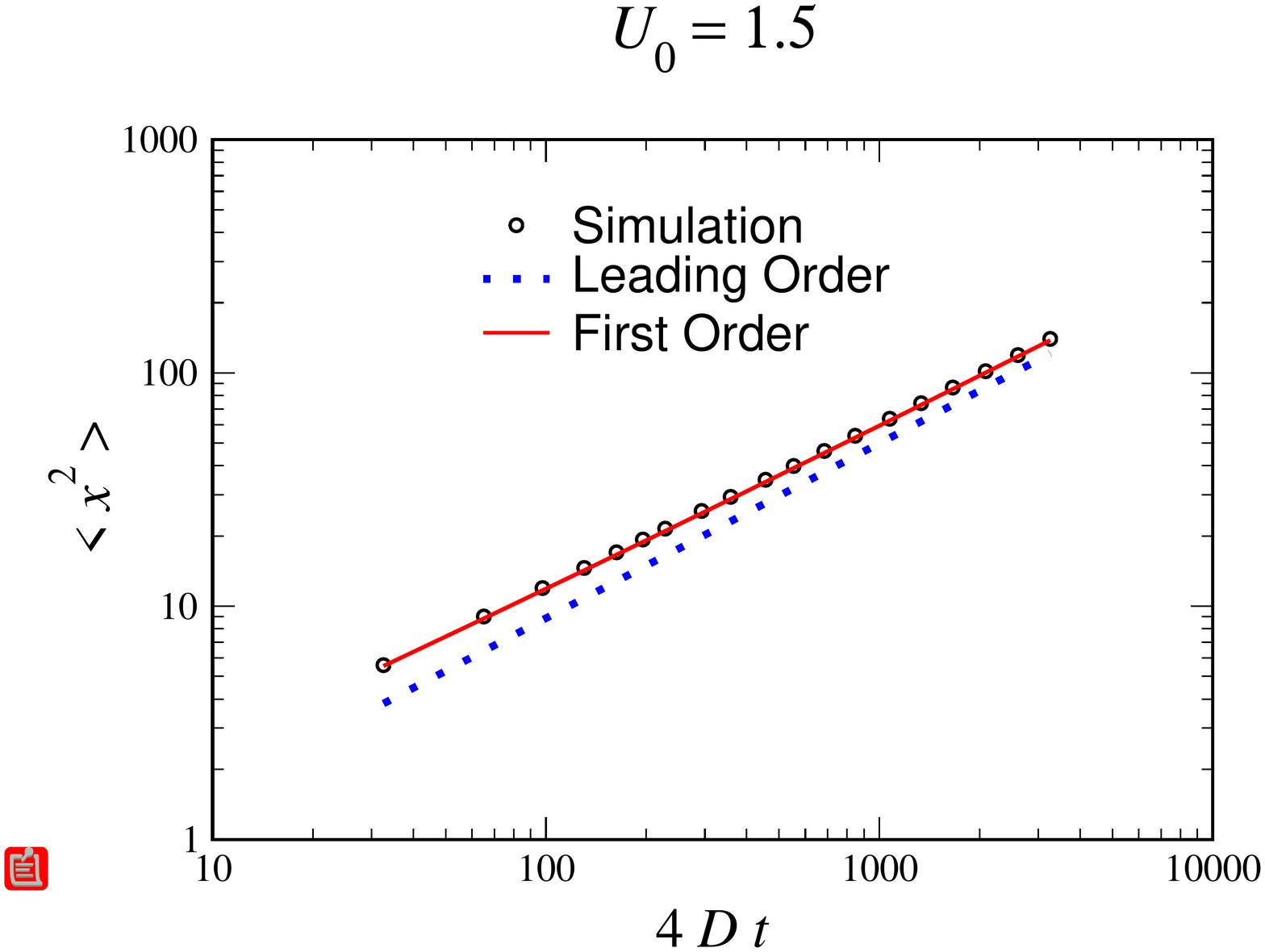}
\caption{$\langle x^2 \rangle$ as a function of time, as measured by simulation of the Fokker-Planck equation for the potential $U(x)=U_0/2\ln(1+x^2)$, with $U_0=2.5$, (left panel), and $U_0=1.5$, (right panel). Also shown are the leading-order power-law result, Eq. (\ref{xsqlead}), and the respective first-order results,
Eq. (\ref{xsqbigalf}) and Eq. (\ref{xsqsmallalf}). Note that for small $t$, $\langle x^2 \rangle \approx 2Dt$, as in simple diffusion~\cite{KesslerPRL}.}
\label{xsqfig}
\end{figure}

In Fig. \ref{xsqfig}, we present $\langle x^2\rangle$ as a function of time, for the cases $U(x)=U_0/2 \ln(1+x^2)$ with $U_0=5/2$,  ($\alpha=7/4$), and $U_0=3/2$, ($\alpha=5/4$).  We also show the first-order corrected predictions, Eqs. (\ref{xsqbigalf}) and (\ref{xsqsmallalf}), respectively, along with the zeroth-order results.  We see in both cases that while the leading-order result is quite inaccurate, due to the slow convergence, the first-order corrected predictions yield excellent agreement.

\subsection{$\alpha <1$ ($U_0 < 1$)}

The calculation for $\alpha<1$ also follows along the same lines as the special case treated in Sec. III.  Returning to Eq. (\ref{Psi}), we now have
\begin{equation}
\Psi_k(x) \approx  \left(\frac{x}{a}\right)^{1/2-\alpha} + k^2 a\left[\left(\frac{x}{a}\right)^{1/2-\alpha} \int_0^x  \frac{s}{2\alpha a} ds - 
\frac{(x/a)^{\alpha+1/2}}{2\alpha } \int_0^x f^2(s)ds \right] \ .
\label{psibigxa}
\end{equation}
The last integral is dominated by its large $x$ behavior, and so grows like $x^{2-2\alpha}$, with a finite, nonuniversal contribution from the finite region.  This finite contribution determines $A_k \sim k^{2-\alpha}$, which shows that $A_k$ is subdominant to $B_k$ for small $k$ for $\alpha<1$.  At this point, the calculation proceeds exactly as in the solvable example, 
 and in fact for large $x$ the result is exactly that  obtained previously for $\alpha<1$, Eq. (\ref{smallalfbigx}). The result is again of scaling form, with a different power-law in time as a prefactor, namely
\begin{equation}
W(x,t)\approx\frac{1}{a}(4Dt/a^2)^{-1/2}{\cal{G}}(x/\sqrt{t}) \ ,
\label{scalingg}
\end{equation}
with
\begin{equation}
 {\cal{G}}(z)=\frac{1}{\Gamma(1-\alpha)} z^{1-2\alpha} e^{-z^2}\ .
 \label{Gz}
 \end{equation}
 Here, as opposed to the case of $\WICD(z,t)$, ${\cal{G}}(z)$ is a true normalized PDF for the scaling variable $z$.  This is connected to the fact that
 the expectation values of $|z|^q$ are time-independent for $\alpha<1$, scaling like normal diffusion.  Although ${\cal{G}}(z)$ is singular at $z=0$, now the
 singularity is integrable.  Again, this singularity points to a breakdown of scaling for small $z$, here it is much less dangerous and does not lead to anomalous behavior.
 
 For $x$ of order 1, we have
\begin{equation}
W(x,t) \approx \frac{e^{-U(x)}}{\Gamma(1-\alpha)a} \left(\frac{4Dt}{a^2}\right)^{\alpha-1} \ ,
\end{equation}
which matches onto the large-$x$ result, Eq. (\ref{smallalfbigx}) and shows that the central region decays away slowly in time.  A uniform approximation is then
\begin{equation}
W(x,t) \approx \frac{e^{-U(x)}}{\Gamma(1-\alpha)a} \left(\frac{4Dt}{a^2}\right)^{\alpha-1} e^{-z^2} \ .
\end{equation}
These results of course reproduce what was found~\cite{KesslerPRL} using the appropriate scaling ansatz, Eq. (\ref{scalingg}).

\begin{figure}
\includegraphics[width=0.7\textwidth]{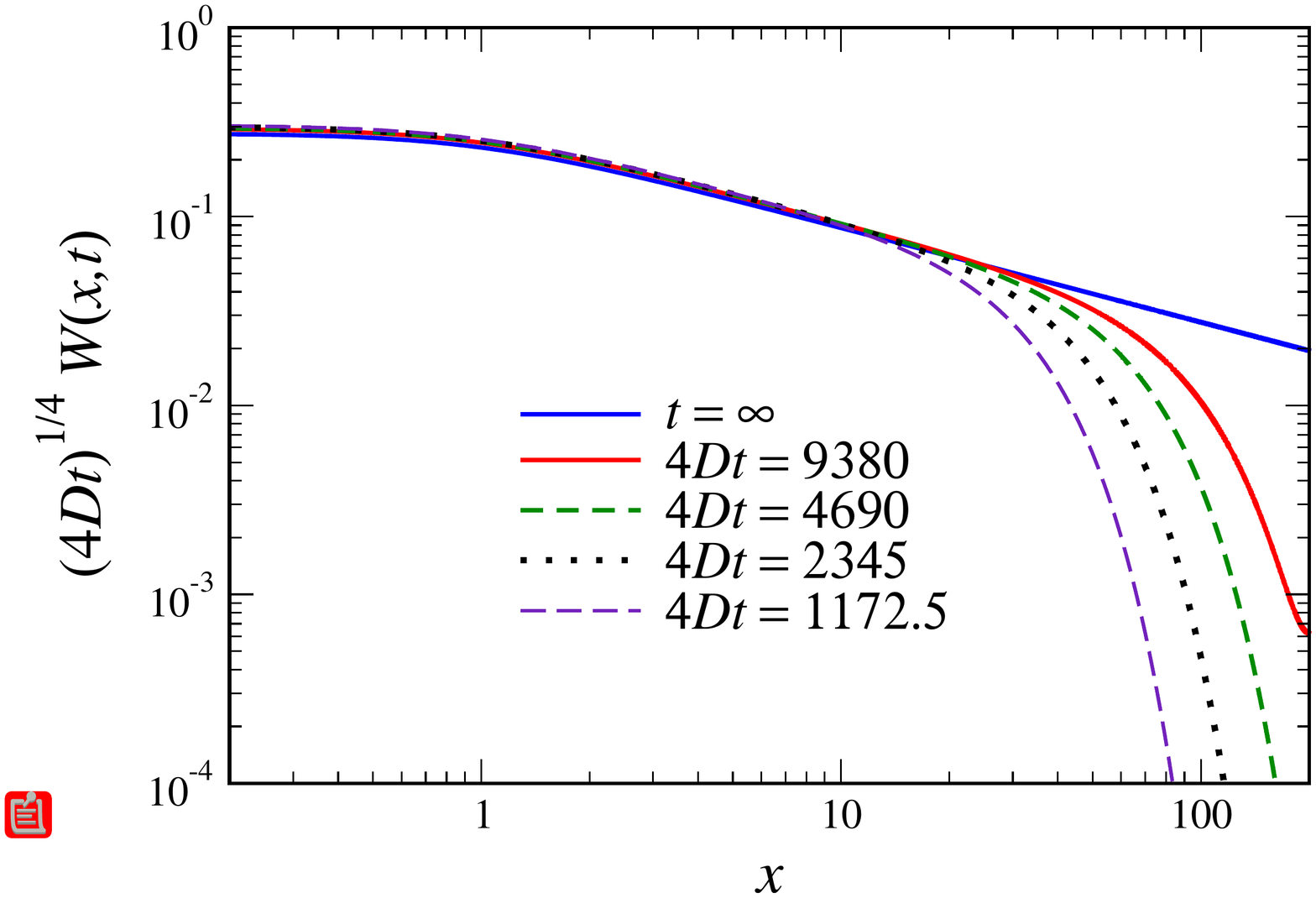}\\
\includegraphics[width=0.7\textwidth]{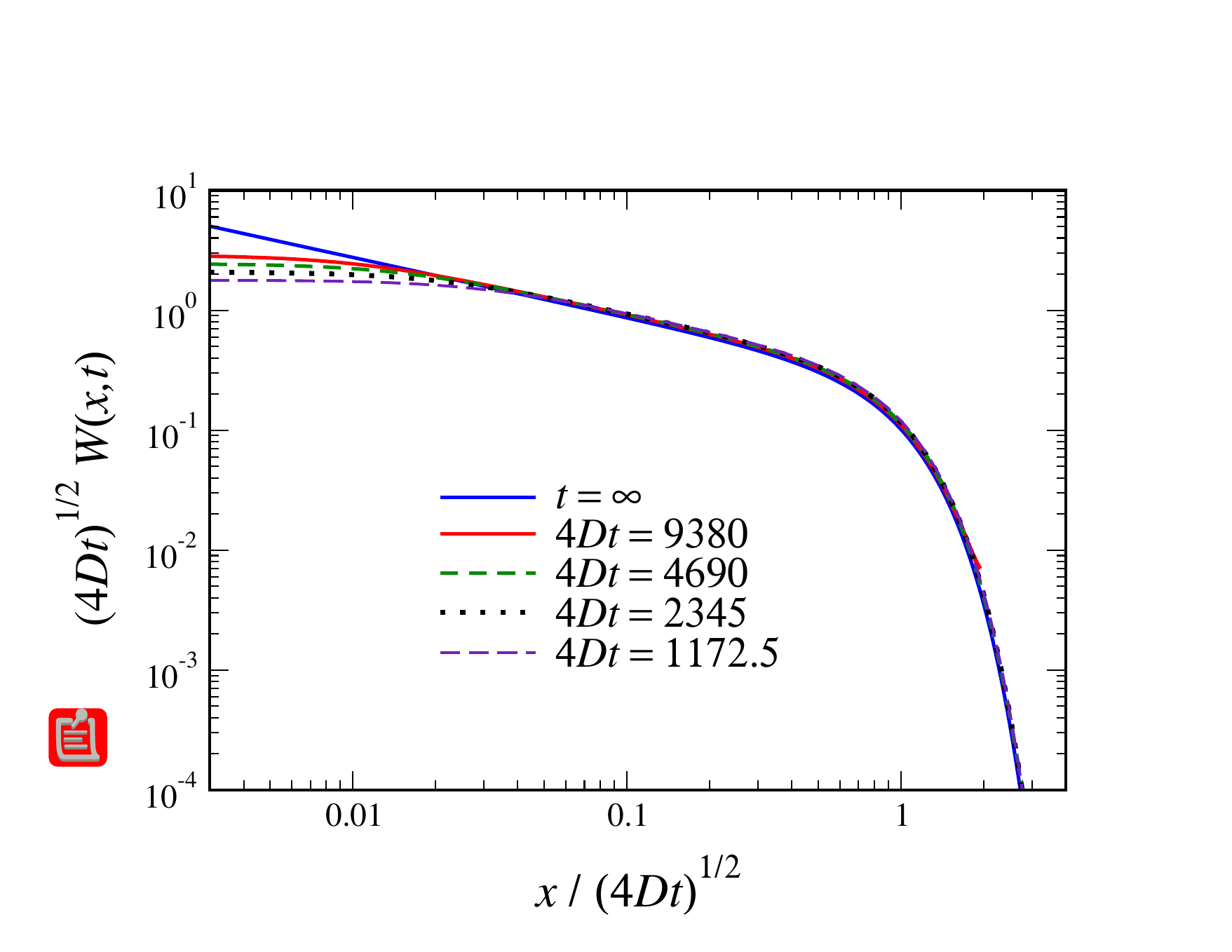}
\caption{(Color online) Top panel: $(4Dt)^{1/4} W(x,t)$ as a function of $x$ for various times for the potential $U(x)=U_0/2\ln(1+x^2)$, $U_0=1/2$, ($\alpha=\nicefrac{3}{4}$), showing the collapse of the central region and the cutoff at large $x$.  The $t=\infty$ curve denotes the infinite time, cutoff-free result, $(1+x^2)^{-{1/4}}/\Gamma(1/4)$.  Bottom panel: The scaling function $(4Dt)^{1/2} W(x,t)$ as a function of $z=x/\sqrt{4Dt}$ for various times, showing the collapse for all but a small region at small $z$, which disappears in the long-time limit, giving the scaling function ${\cal{G}}(z)$, Eq. (\ref{Gz}).}
\label{fig3}
\end{figure}

In the top panel of Fig. \ref{fig3} we present $(4Dt)^{1-\alpha}W(x,t)$ as a function of $x$ for the potential $U(x)=\nicefrac{1}{4}\ln(1+x^2)$, so that $\alpha=\nicefrac{3}{4}$, $a=1$.  Also shown is the asymptotic result $e^{-U(x)}/\Gamma(1-\alpha) = (1+x^2)^{-\nicefrac{1}{4}}/\Gamma(\nicefrac{1}{4})$, denoted as
$t=\infty$.  The collapse of the PDF to the nonnormalizable Boltzmann distribution in the central region is clear, along with the cutoff at  $x\sim{\cal{O}}((Dt)^{1/2})$.  In the bottom panel, the scaling collapse $(4Dt)^{1/2} W(x,t)$ as a function of $z$ is shown.  This collapse of course breaks down for small $z$, i.e, for $x$ of order unity. 

\begin{figure}
\includegraphics[width=0.7\textwidth]{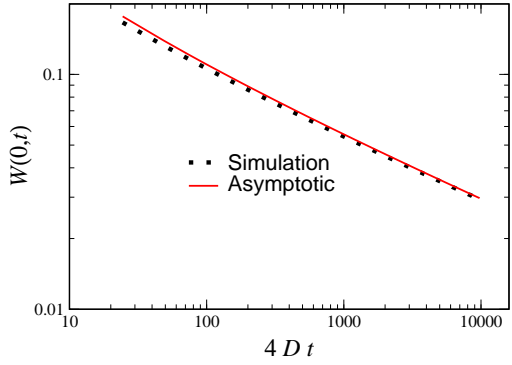}
\caption{$W(0,t)$ versus $4Dt$ for the case presented in Fig. \ref{fig3} together with the asymptotic approximation for $1/Z(t)$, with $Z(t)$ given by
Eq. (\ref{Z1}).}
\label{fig4}
\end{figure}

As in the case of $\alpha>1$, there is a algebraically decaying correction to the normalization, coming from the breakdown of the scaling solution at small
$z$, which itself is decaying in time.  If we define $\tilde{Z}(t)\equiv \int_{-\infty}^\infty dx\, e^{-U(x)} e^{-z^2}$, we find
\begin{eqnarray}
\tilde{Z}(t) &\approx& 2\int_{0}^\infty dx\, (x/a)^{1-2\alpha} e^{-z^2} - 2\int_0^\infty dx\, \left[(x/a)^{1-2\alpha} - e^{-U(x)}\right] \nonumber\\
&=& \Gamma(1-\alpha) a \left(\frac{4Dt}{a^2}\right)^{1-\alpha} - 2\int_0^\infty dx\, \left[(x/a)^{1-2\alpha} - e^{-U(x)}\right] \ .
\label{Ztw}
\end{eqnarray}
The second integral of course depends on the specific form of $U(x)$.  For example, if $U(x)=U_0/2\ln(1+x^2)$, we get
\begin{eqnarray}
\int_0^\infty dx\, \left[(x/a)^{1-2\alpha} - (1+(x/a)^2)^{1/2-\alpha}\right] &=&(\alpha-1/2)\int_0^\infty dx\, \int_0^1 ds\, (s+(x/a)^2)^{-1/2-\alpha}\nonumber\\
&=& a(\alpha-1/2) \int_0^1 ds\, \frac{\sqrt{\pi}\Gamma(\alpha)}{2\Gamma(\alpha+1/2)} s^{-\alpha} \nonumber\\
&=& \frac{a\sqrt{\pi}\Gamma(\alpha)}{2\Gamma(\alpha-1/2)(1-\alpha)}\ .
\end{eqnarray}
Thus, for this family of potentials,
\begin{equation}
\tilde{Z}(t) = \Gamma(1-\alpha) a \left(\frac{4Dt}{a^2}\right)^{1-\alpha} -   \frac{a\sqrt{\pi}}{(1-\alpha)\Gamma(\alpha-\nicefrac{1}{2})} \Gamma(\alpha)  \ .
\label{Z1}
\end{equation}
In Fig. \ref{fig4}, we show $W(0,t)$ as a function of time, together with the analytic prediction $W(0,t)\approx 1/\tilde{Z}(t)$.

The calculation of the second moment  to leading order is straightforward~\cite{KesslerPRL}:
\begin{eqnarray}
\langle x^2 \rangle &\approx& \frac{2}{\Gamma(1-\alpha)a}\int_0^\infty dx\, x^2 z^{1-2\alpha} \left(\frac{4Dt}{a^2}\right)^{-1/2}e^{-z^2} \nonumber\\
&=& 4D(1-\alpha)t \ .
\end{eqnarray}
so that in this regime the growth of the second moment is pure diffusive in nature, with a renormalized diffusion constant.

\begin{figure}
\includegraphics[width=0.7\textwidth]{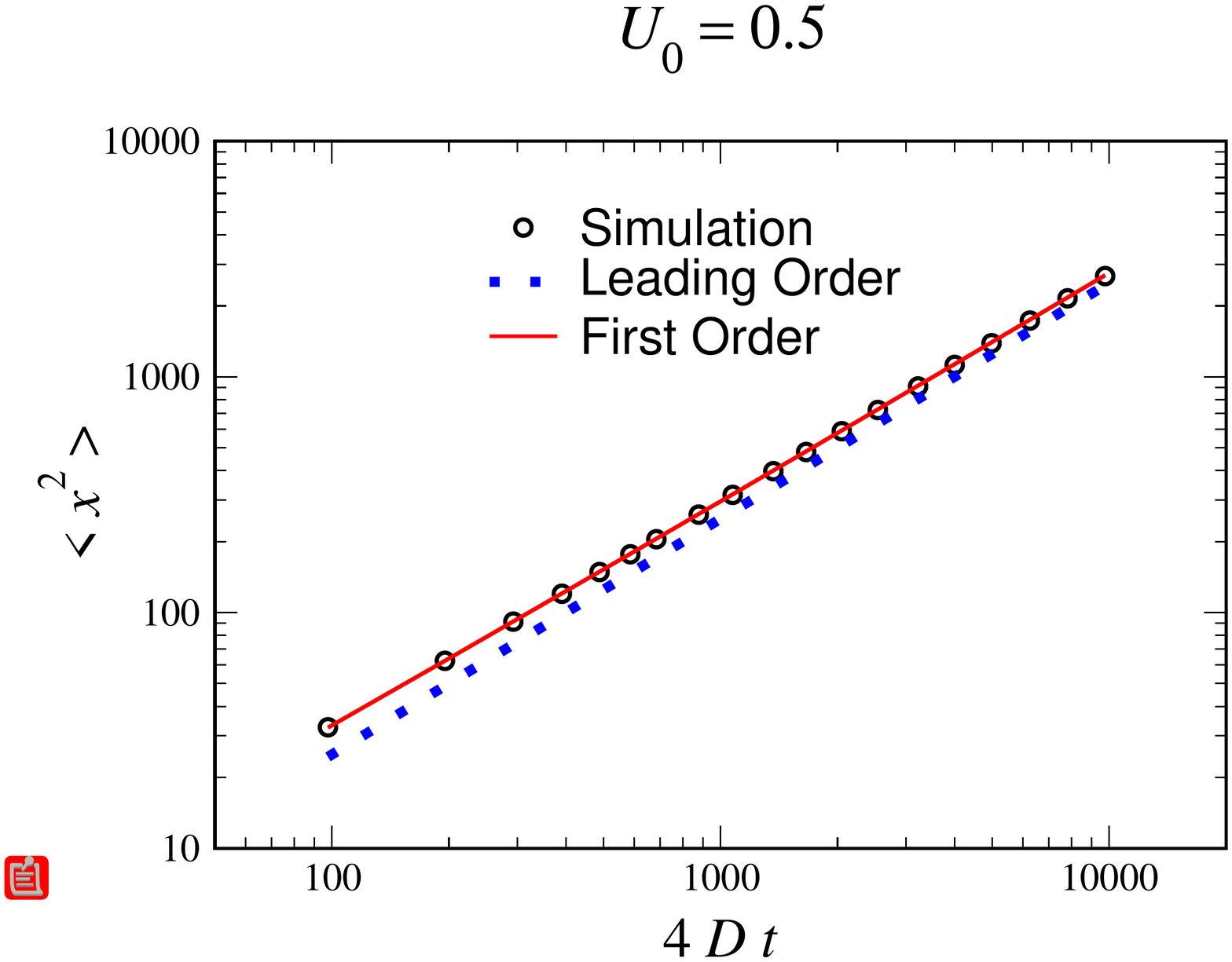}
\caption{$\langle x^2 \rangle$ as a function of $t$, as measured by simulation of the Fokker-Planck equation for the case presented in Fig. \ref{fig3}. Also shown are the leading-order power-law result, Eq. (\ref{xsqlead}), and the  first-order result,
Eq. (\ref{xsqlowalf}).}
\label{xsqfig2}
\end{figure}

As in the case $\alpha>1$, the corrections to this result can be relatively large, as they decay quite slowly in time.  This is apparent from Eq. (\ref{Z1}), where the correction term is of relative order $t^{\alpha -1}$.  While in the central region, the correction is just a change in normalization, in the outer region, the first-order correction has a different functional form.  We assume the correction to $W$ is of the form
\begin{equation}
W_1(x,t) \approx (4Dt/a^2)^{\alpha-3/2} {\cal{G}}_1(z) ; \qquad\qquad {\cal{G}}_1 \sim z^{1-2\alpha}\quad \textrm{for}\quad z\ll 1
\end{equation}
which implies that ${\cal{G}}_1$ satisfies
\begin{equation}
{\cal{G}}_1'' + \left(\frac{2\alpha-1}{z} + 2z\right) {\cal{G}}_1' - \left(\frac{2\alpha-1}{z^2} + 4\alpha - 6\right) {\cal{G}}_1 = 0 \ .
\end{equation}
The solution which matches onto the central region correction induced by the correction to $\tilde{Z}(t)$, Eq. (\ref{Ztw}), is
\begin{equation}
{\cal{G}}_1(z) = C z e^{-z^2} U(2\alpha-1,1+\alpha,z^2); \qquad\qquad C=2{\cal{I}}\frac{\Gamma(2\alpha-1)}{ \Gamma(\alpha)\Gamma^2(1-\alpha)a^2} \ ,
\end{equation}
where
\begin{equation}
{\cal{I}} \equiv \int_0^\infty dx\, \left[(x/a)^{1-2\alpha} - e^{-U(x)}\right] \ .
\end{equation}
$W_1$ induces a correction in $\langle x^2 \rangle$:
\begin{eqnarray}
\langle x^2 \rangle &\approx& 4D(1-\alpha)t + 2\left(\frac{4Dt}{a^2}\right)^{\alpha-3/2}\int_0^\infty dx\, x^2 C z e^{-z^2} U(2\alpha-1,1+\alpha,z^2) \nonumber\\
&=& 4D(1-\alpha)t + \left(\frac{4Dt}{a^2}\right)^{\alpha} Ca^3 \int_0^\infty ds\,  s e^{-s} U(2\alpha-1,1+\alpha,s) \nonumber\\
\nonumber\\
&=& 4D(1-\alpha)t + \left(\frac{4Dt}{a^2}\right)^{\alpha}Ca^3 \frac{\Gamma(2-\alpha)}{\Gamma(1+\alpha)} \nonumber\\
&=& 4D(1-\alpha)t + \left(\frac{4Dt}{a^2}\right)^{\alpha}a \frac{2{\cal{I}}\Gamma(2\alpha-1)\Gamma(2-\alpha)}{\Gamma(1+\alpha)\Gamma(\alpha)\Gamma^2(1-\alpha)} \ .
\label{xsqlowalf}
\end{eqnarray}
In Figure \ref{xsqfig2}, we present $\langle x^2 \rangle$ for the case $U(x)=U_0/2\ln(1+x^2)$, $U_0=\nicefrac{1}{2}$, ($\alpha=\nicefrac{3}{4}$), together with the leading and first order predictions.  Again, we see that the correction term is important even when the measurement time is quite large, e.g., $4Dt = 10^4$.

 \section{Asymmetric Initial Conditions}
 If we start with an asymmetric initial condition, (a delta function at $x=x_0$, for example), the first moment is not guaranteed to vanish except at infinite time.
 For a ``normal" potential, with a discrete spectrum, the first moment decays exponentially.
 In the pure diffusion case, $U_0=0$, however, the system never forgets the initial asymmetry and the first moment is preserved for all time.  It is interesting
 then to calculate the decay of the first moment.
 The first order of business is to compute the odd eigenfunctions. The outer solution for the odd states, which we denote by $\phi_k(x)$, is again a linear combination of Bessel functions:
 \begin{equation}
\phi_k(x)= M_k \sqrt{\frac{x}{a}} \left[C_k J_\alpha(kx) + D_k J_{-\alpha}(kx)\right] \ .
\end{equation}
The inner, i.e., $x\sim {\cal{O}}(1)$, odd solution we write as $\phi_k = M_k \Phi_k$, with
\begin{equation}
\Phi_k = g(x) - k^2a \left[g(x)\int_0^x g(s)f(s)ds - f(x)\int_0^x g^2(s)ds\right]
\label{Phi}
\end{equation}
in analogy with Eq. (\ref{Psi}).
The leading order behavior of $\Phi_k$ for large $x$ matches with the small $kx$ behavior of $J_\alpha(kx)$ in our matching region $1 \ll x \ll 1/k$, so that 
\begin{equation}
C_k \approx (ka)^{-\alpha} {\cal{C}}; \qquad\qquad {\cal{C}} \approx 2^{\alpha-1} \Gamma(\alpha) \ .
\end{equation}
The leading order behavior of $D_k$ comes from the constant piece in the large $x$ asymptotics of the last integral, and is nonuniversal, but by matching the $k$ dependence of $D_k J_{-\alpha}$ to the $k^2$ coefficient of $f(x)$, we get that $D_k \sim k^{2+\alpha}$, and
so $C_k$ dominates for all $\alpha$.  The normalization $M_k$ is then
\begin{equation}
(M_k)^2\approx \frac{\pi}{2{\cal{C}}^2L} (ka)^{2\alpha+1} \ .
\end{equation}
 Thus,  we get, for a $\delta(x-x_0)$ ($x_0 \ne 0$) initial condition, the contribution of the odd modes:
\begin{equation}
W^\textit{\scriptsize{odd}}(x,t) \approx \frac{e^{-U(x)/2+U(x_0)/2}}{2{\cal{C}}^2} \int_0^\infty dk\, (ka)^{\alpha+1}\Phi_k(x_0) \Phi_k(x) e^{-Dk^2 t} \ .
\end{equation}
For large  $t$, the relevant $k \sim t^{-1/2}$, and so $kx_0$ is arbitrarily small and $\Phi_k(x_0)$ can be approximated by $g(x_0)$. Then,  for large $x$, 
\begin{eqnarray}
W^\textit{\scriptsize{odd}} &\approx& \frac{e^{-U(x)/2+U(x_0)/2}g(x_0)}{2{\cal{C}}}\sqrt{\frac{x}{a}} \int_0^\infty dk (ka)^{\alpha+1} J_\alpha(kx) e^{-Dk^2 t}\nonumber\\
&=&\frac{e^{-U(x)/2+U(x_0)/2}g(x_0)}{4a{\cal{C}}}\sqrt{\frac{x}{a}}\left(\frac{Dt}{a^2}\right)^{-1-\alpha/2} z^{\alpha}e^{-z^2} \nonumber\\
&=& \frac{2e^{U(x_0)/2}g(x_0)}{a\Gamma(\alpha)}\left(\frac{4Dt}{a^2}\right)^{-1/2-\alpha}z e^{-z^2} \ .
\end{eqnarray}
A more pleasing way of writing this is in term of the unnormalized even and odd solutions of the Fokker-Planck equation, 
\begin{equation}
X(x) \equiv e^{U(x)};  \qquad\qquad Y(x) \equiv \frac{1}{a}e^{-U(x)}\int_0^x dy\, e^{U(y)} \ ,
\end{equation}
satisfying $X(0)=1$, $Y'(0)=1/a$ and $Y(x) \sim x/(2\alpha a)$ for large $x$.
Then
\begin{equation}
W^\textit{\scriptsize{odd}} \approx \frac{2Y(x_0)}{aX(x_0)\Gamma(\alpha)}\left(\frac{4Dt}{a^2}\right)^{-\nicefrac{1}{2}-\alpha} z e^{-z^2} \ .
\end{equation}
This odd solution has the scaling structure $W^\textit{\scriptsize{odd}} \sim t^{\beta} {\cal{F}}_\textit{\scriptsize{odd}}(z)$, $\beta=-\nicefrac{1}{2}-\alpha$. Whereas for the even solution, we could guess the exponent $\beta$, since we could demand that it match the time-independent leading order solution, here
there seems to be no way to guess $\beta$ without using our eigenvalue expansion.   Plugging the scaling ansatz into the Fokker-Planck equation gives the scaling equation for ${\cal{F}}_\textit{\scriptsize{odd}}$:
\begin{equation}
0 = {\cal{F}}_\textit{\scriptsize{odd}}'' +  \left(\frac{2\alpha-1}{z} + 2z\right){\cal{F}}_\textit{\scriptsize{odd}}' + \left(2+4\alpha - \frac{2\alpha-1}{z^2}\right){\cal{F}}_\textit{\scriptsize{odd}} \ ,
\end{equation}
for which ${\cal{F}}_\textit{\scriptsize{odd}}(z)=ze^{-z^2}$ is clearly a solution.
For $x$'s of order 1,
\begin{eqnarray}
W^\textit{\scriptsize{odd}}(x,t) &=& \frac{e^{-U(x)/2+U(x_0)/2}}{2{\cal{C}}^2} \int_0^\infty g(x_0)g(x) (ka)^{2\alpha+1} e^{-Dk^2 t} \nonumber\\
&=& \frac{4\alpha ZY_0(x_0)Y_0(x)}{X(x_0)\Gamma(\alpha)}  \left(\frac{4Dt}{a^2}\right)^{-1-\alpha} \ ,
\end{eqnarray}
so that for $1 \ll x \ll \sqrt{4Dt}$, this approaches our previous result.  A uniform approximation for the $W^\textit{\scriptsize{odd}}$ is then
\begin{equation}
W^\textit{\scriptsize{odd}} \approx \frac{4\alpha Y_0(x_0)Y_0(x)}{aX(x_0)\Gamma(\alpha)}  \left(\frac{4Dt}{a^2}\right)^{-1-\alpha} e^{-z^2} \ .
\end{equation}
We see that for all $\alpha$ this decays in time as a power law which is faster than the decay of the even part.  Furthermore, it is $x_0$ dependent, as opposed to the even part which to leading order is independent of $x_0$.

\begin{figure}
\includegraphics[width=0.7\textwidth]{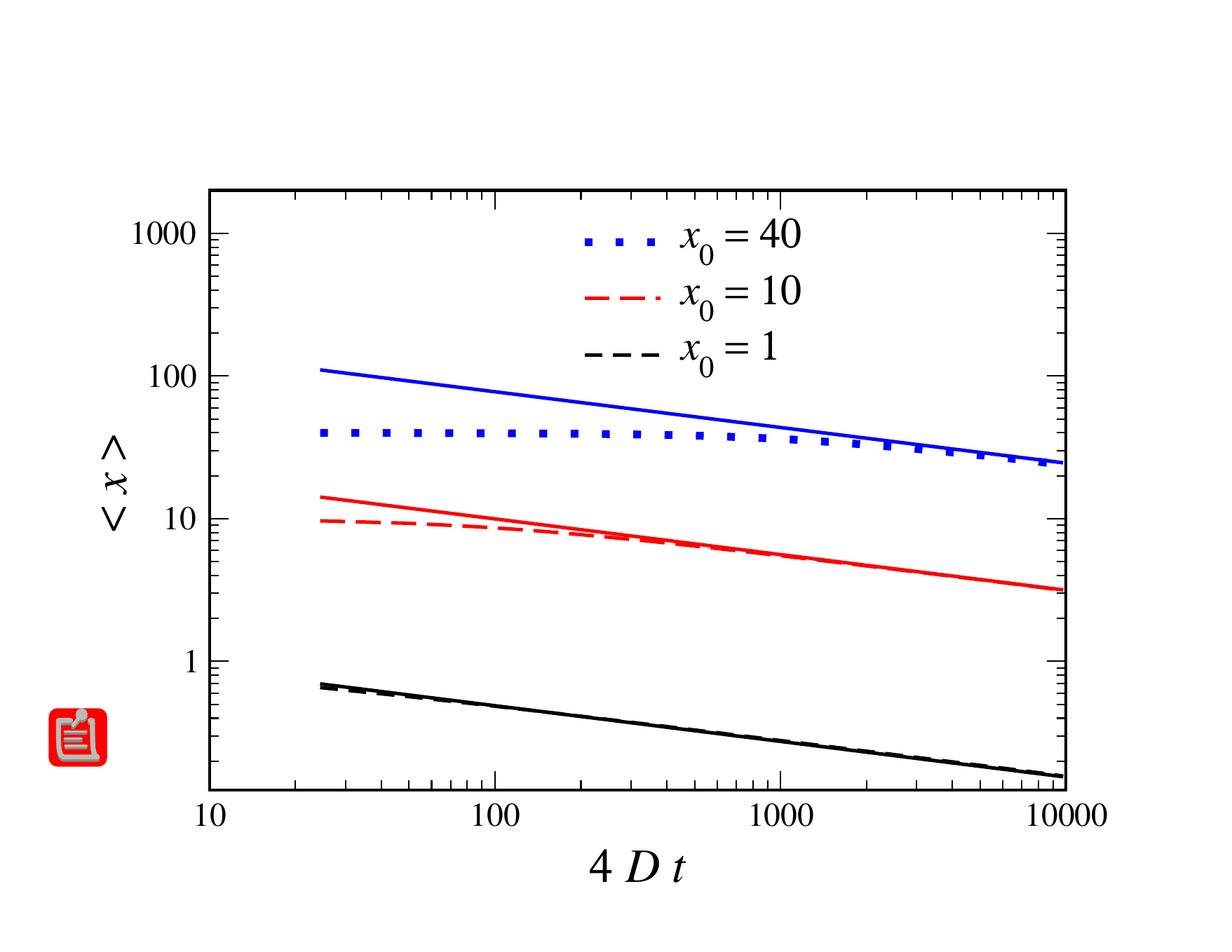}
\caption{$\langle x \rangle$ versus $4Dt$ for the potential $U(x)=\nicefrac{1}{4}\ln(1+x^2)$, for the cases $x_0=1$, (blue dotted line), $10$, (red long dashed line), and $40$, (purple short dashed line), together with the asymptotic approximations, Eq. (\ref{xbar}), shown as the corresponding solid lines.}
\label{fig5}
\end{figure}

The calculation of the first moment is then straightforward, since it is dominated by the large $x$ behavior:
\begin{eqnarray}
\langle x \rangle &\approx& \frac{4Y(x_0)}{aX(x_0)\Gamma(\alpha)}\left(\frac{4Dt}{a^2}\right)^{-\nicefrac{1}{2}-\alpha} \int_0^\infty dx\, xz e^{-z^2}\nonumber\\
&=& \frac{a\sqrt{\pi}Y(x_0)}{X(x_0)\Gamma(\alpha)}  \left(\frac{4Dt}{a^2}\right)^{\nicefrac{1}{2}-\alpha} \ .
\label{xbar}
\end{eqnarray}
In particular, when $U_0=0$, i.e., $\alpha=1/2$, the first moment does not decay at all.  Since $Y(x_0)$ vanishes at $x_0=0$ and grows for large $x_0$
as $x_0$, the first moment grows as $x_0^{2\alpha}$, so starting with a fairly large $x_0$ should make measurement of this decay much easier.  We see this in Fig. \ref{fig5}, where we present $\langle x\rangle$ as a function of time, for $x_0=1$, $10$, and $40$, for the potential $U(x)=\nicefrac{1}{4}\ln(1+x^2)$.  We see that even though the time for the asymptotic decay to set in gets much larger as $x_0$ increases, the value of the first moment at the onset of the asymptotic scaling regime is still much larger for the larger $x_0$.

\section{Conclusions}
We have explicitly constructed the probability density function for diffusing Brownian particles in an attractive logarithmic-type potential, $V(x)\sim V_0 \ln x$, $V_0>0$.  The system exhibits different modes of behavior as the strength of the potential is varied.  For very weak potentials, $U_0\equiv V_0/k_B T<1$, the particles diffuse as $t^{1/2}$ (to leading order) with a renormalized diffusion constant, and with algebraic corrections. The Boltzmann distribution in this case is nonnormalizable, yet nevertheless the PDF, for $x \ll \sqrt{4Dt}$, is
prorportional to the Boltzmann density, with a algebraically decaying prefactor.  For $1<U_0<3$, we see that both the normalizable Boltzmann distribution and the infinite covariant density are necessary for a full description of the leading-order dynamics.  
The probability density approaches the equilibrium distribution for $x$'s smaller than ${\cal{O}}(\sqrt{Dt})$ in magnitude.  For larger $x$'s, the PDF decays as a Gaussian.  The second, and higher, moments of the distribution are sensitive to the cutoff, grow algebraically with time, and can be calculated from the infinite covariant density. For larger $U_0$, the second moment saturates at its equilibrium value, but higher moments are sensitive to the cutoff and grow as a power-law in time.  These cutoff-sensitive moments are calculable via the infinite covariant density, which is the scaling limit of the PDF. Thus, in all cases, the Boltzmann distribution describes the $x\ll \sqrt{4Dt}$ regime, and a scaling function describes the PDF at larger $x$. The leading-order calculation of 
the moments is subject to corrections which decay quite slowly in time (i.e. with a power less than one-half), and so for reasonably large times, the leading-order prediction is quantitatively poor.  Upon adding the first-order correction, the prediction of the 0th and second moments is seen to be quantitatively accurate.  The first moment of the distribution, unlike for pure diffusion where it stays constant, decays, albeit very slowly, as $t^{-U_0/2}$.

\begin{acknowledgements}
This  work  was supported by the  Israel Science  Foundation, the Emmy Noether Program of the DFG (contract No LU1382/1-1) and the cluster of excellence Nanosystems Initiative Munich.
\end{acknowledgements}

\bibliography{bibfile}

\end{document}